\documentclass[manuscript]{aastex61}
\usepackage{subfigure}
\newcommand{\Chandra}{{\it Chandra}}
\newcommand{\Suzaku}{{\it Suzaku}}
\newcommand{\XMM}{{\it XMM-Newton}}
\newcommand{\ROSAT}{{\it ROSAT}}

\usepackage{color}
\usepackage{physics}
\usepackage{bm}
\usepackage{booktabs}

\submitjournal{ApJ}

\shorttitle{}
\shortauthors{Zhu et al.}

\begin{document}

\title{A Study of Gas Entropy Profiles of 47 Galaxy Clusters and Groups Out to the Virial Radius}

\author{Zhenghao Zhu}
\affiliation{School of Physics and Astronomy, Shanghai Jiao Tong University, 800 Dongchuan Road, Minhang, Shanghai 200240, China}
\correspondingauthor{Haiguang Xu}
\author{Haiguang Xu}
\affiliation{School of Physics and Astronomy, Shanghai Jiao Tong University, 800 Dongchuan Road, Minhang, Shanghai 200240, China}
\affiliation{IFSA Collaborative Innovation Center, Shanghai Jiao Tong University, 800 Dongchuan Road, Shanghai 200240, China}
\email{hgxu@sjtu.edu.cn}
\author{Dan Hu}
\affiliation{School of Physics and Astronomy, Shanghai Jiao Tong University, 800 Dongchuan Road, Minhang, Shanghai 200240, China}
\author{Chenxi Shan}
\affiliation{School of Physics and Astronomy, Shanghai Jiao Tong University, 800 Dongchuan Road, Minhang, Shanghai 200240, China}
\author{Yongkai Zhu}
\affiliation{School of Physics and Astronomy, Shanghai Jiao Tong University, 800 Dongchuan Road, Minhang, Shanghai 200240, China}
\author{Shida Fan}
\affiliation{School of Physics and Astronomy, Shanghai Jiao Tong University, 800 Dongchuan Road, Minhang, Shanghai 200240, China}
\author{Yuanyuan Zhao}
\affiliation{School of Physics and Astronomy, Shanghai Jiao Tong University, 800 Dongchuan Road, Minhang, Shanghai 200240, China}
\author{Liyi Gu}
\affiliation{RIKEN High Energy Astrophysics Laboratory, 2-1 Hirosawa, Wako, Saitama 351-0198, Japan}
\affiliation{SRON Netherlands Institute for Space Research, Sorbonnelaan 2, 3584 CA Utrecht, the Netherlands}
\author{Xiang-Ping Wu}
\affiliation{National Astronomical Observatories,
	Chinese Academy of Sciences,
	20A Datun Road, Beijing 100012, China}



\begin{abstract}

Some observations such as those presented in Walker et al. show that the observed entropy profiles of the intra-cluster medium (ICM) deviate from the power-law prediction of adiabatic simulations. This implies that non-gravitational processes, which are absent in the simulations, may be important in the evolution of the ICM, and by quantifying the deviation, we may be able to estimate the feedback energy in the ICM and use it as a probe of the non-gravitational processes. 
To address this issue we calculate the ICM entropy profiles in a sample of 47 galaxy clusters and groups, which have been observed out to at least  $\sim r_{500}$ with \Chandra{}, \XMM{} and/or \Suzaku{}, by constructing a physical model to incorporate the effects of both gravity and non-gravitational processes to fit the observed gas temperature and surface brightness profiles via Bayesian statistics. After carefully evaluating the effects of systematic errors, we find that the gas entropy profiles derived with best-fit results of our model are consistent with the simulation-predicted power-law profile near the virial radius, while the flattened profiles reported previously can be explained by introducing the gas clumping effect, the existence of which is confirmed in 19 luminous targets in our sample.
We calculate the total feedback energy per particle and find that it decreases from $\sim 10$ keV at the center to about zero at $\sim 0.35$$r_{200}$ and is consistent with zero outside $\sim 0.35$$r_{200}$, implying the upper limit of the feedback efficiency $\sim 0.02$ for the super-massive black holes hosted in the brightest cluster galaxies. 
\end{abstract}




\section{Introduction} \label{sec:intro}
In terms of the hierarchical structure formation scenario \citep{ps74} the primordial density fluctuations grow into galaxies, which evolve into proto-clusters and then more massive systems via accretion of surrounding materials and mergers with other clusters \citep{voit05b,TNG18}.
Although the whole process is expected to be dominated by gravity, the non-gravitational processes including the feedback of active galactic nuclei (AGN), which has become a very active topic in the recent decade (see \citealt{fabian12} for a review), radiation of the intra-cluster medium (ICM), conduction within the ICM, a possible pre-heating at the early stage of the cluster formation, etc., may also have chances to play important roles (e.g., \citealp{vazza11}).  
In fact, considerable or relatively minor heat-exchanging happens in all these processes. Therefore the thermodynamical properties of the ICM change accordingly, depending on the time and space scales considered \citep{iqbal17}. For example, AGNs can break the hydrostatic equilibrium via heating the ICM in the core regions with the energetic outflows \citep{vazza13}, and the radiative cooling of the gas tends to produce a cooler but brighter cluster core, which often appears as a so-called cool core \citep{hudson10}. Thus the theoretically predicted mass-luminosity and mass-temperature relations are broken \citep{kaiser86}. 
Therefore the details of non-gravitational processes must be carefully treated while studying the structures of the ICM and their evolutionary histories (e.g., \citealt{voit02}; \citealt{planelles13}; \citealt{lovell18}).  

Among various gas properties the entropy $K$, usually defined as $T\times n_{\rm e}^{-2/3}$ in astrophysical literature \citep{voit05b}, where $T$ and $n_{\rm e}$ are the temperature and the electron density of the ICM, respectively, is one of the probes most sensitive to the non-gravitational processes, because it is the logarithm of the entropy $S$ in thermodynamics, which is directly related to the net change of heat energy $dQ$ via $dQ=TdS$ (see Section \ref{sec:model} for detail). By comparing the observed gas entropy profiles with those predicted under the assumption of pure gravity, we may be able to evaluate the contribution of non-gravitational processes and estimate the corresponding net heat supply in the ICM (e.g., \citealt{cnm12}; \citealt{iqbal17}).

Results of non-radiative (i.e., no feedback or cooling is considered) hydrodynamical simulations show that outside the cluster core the ICM entropy profile scales as $\propto r^{1.1}$ (e.g., \citealt{voit02}; \citealt{voit05b}). Studies of, e.g., \citet{su15}, \citet{tcher16}, etc., do show that the observed gas entropy profiles are consistent with the power-law profile out to $r_{200}$. On the other hand, however, studies based on many X-ray observations performed with \Chandra{}, \XMM{} and \Suzaku{} (e.g., \citealt{bautz09}; \citealt{kawa10}; \citealt{akama11}; \citealt{ichikawa13}; \citealt{ghirar18}) indicate that near the virial radius the observed gas entropy profiles differ more or less from the power-law prediction. \citet{walker12a} (hereafter W12a) studied a sample of 11 clusters ($z<0.25$) probed out to $\sim r_{200}$\footnote{$r_\delta$ is the radius within which the mean mass density is $\delta$ times the critical density of the local universe.} and found that most of the observed entropy profiles follow a universal shape that starts to flatten at around $r_{500}$. In order to explain these confusing inconsistent results, \citet{ghirar18} have attempted to attribute the flattening of the gas entropy profiles detected in their sample of 12 clusters to the fact that the gas clumping effect is not taken into account in modeling the X-ray surface brightness. In fact, if the actual gas distribution is clumpy, the assumption of uniform distribution of gas density in the model is found to result in a higher average gas density and then a lower entropy (e.g., \citealp{ron06}; \citealp{vazza11}), causing the entropy profile flatter than the power-law prediction.
Besides the clumping effect, either the dynamical non-equilibrium of ICM caused by bulk motions or turbulence (\citealp{okabe14}; \citealp{khatri16}), or the unbalance thermodynamical state between the electron or ion populations \citep{hoshino10}, or the adiabatic expansion caused by the weakening of the accretion processes in a relaxed cluster (\citealp{lapi10}) has also been proposed to explain the flattening of the entropy profiles near the virial radius. The reason for the inconsistency remains controversial up to now.

In order to solve this interesting problem, a large sample analyzed with a high signal to noise ratio (S/N) and a sufficiently good spatial resolution is apparently needed. In this work we built a sample containing 47 galaxy clusters and groups that have been observed out to at least $\sim r_{500}$, and applied a revised thermodynamical ICM model (RTI model), which is improved from the analytic model presented in \citet{zhu16} (hereafter Z16), to describe the observed gas temperature, surface brightness and total mass distributions. By using the best-fit parameters to investigate the gas entropy distributions, which are evaluated with the uncertainties caused by different systematic effects primarily in instrument calibrations and in the modeling of the plasma emission, we conclude that the gas entropy profiles near the virial radius are consistent with the power-law prediction within the $68\%$ confidence range. We also calculate the feedback energy using the derived gas entropy profiles and obtain a moderate feedback efficiency of $\sim 0.02$ for the super-massive black holes (SMBHs) hosted in the brightest cluster galaxies (BCGs).

This paper is organized as follows. In Section 2 we describe the sample selection criteria. In Section 3 we describe the data analysis procedure for \Chandra{} observations, which covers the $\sim 0.3$  $r_{500}$ regions for all sample targets. Meanwhile we search in literature for works based on either \Chandra{}, or \XMM{}, or \Suzaku{} observations that cover out to at least $r_{500}$, and quote the observed gas temperature and X-ray surface brightness presented therein. These profiles, together with those obtained in our \Chandra{} analysis for the inner $0.3$ $r_{500}$ will be fed into our model as observational constraints. In Section 4 we describe our model and use it to fit the observed gas temperature and X-ray surface brightness profiles prepared in Section 3, and use the best-fit results to calculate the gas entropy profiles. Our results are discussed in Section 5 and summarized in Section 6. 
Throughout this work we adopt a flat $\Lambda$CDM cosmology with $\Omega_m=0.27$, $\Omega_\lambda=0.73$, and the Hubble constant $H_0=71$ km ${\rm s^{-1}}$ ${\rm Mpc^{-1}}$. Unless otherwise stated we use the solar abundance standards of \citet{grevesse98} and quote errors at the $68\%$ confidence level.

\section{Sample construction}
In order to characterize the entropy profiles accurately and precisely with sufficient spatial resolution, we construct our sample from the clusters and groups satisfying the following criteria. (1) The X-ray surface brightness and gas temperature of the target should have been measured out to at least $\sim r_{500}$ with either \Chandra{}, or \XMM{}, or \Suzaku{} in previous studies (Table \ref{sample}) with available gas temperature and X-ray surface brightness data that can be quoted and used in the RTI model fitting. 
(2) Because the extended, irregular, and energy-dependent point spread functions (PSFs) of \XMM{} and \Suzaku{} (e.g., \citealt{snowden08}; \citealt{sugizaki09}) have a non-negligible effect on the analysis of the observed data, especially in central regions, we download and analyze the corresponding \Chandra{} data for the inner $0.3$ $r_{500}$ to better constrain the RTI model fitting. 
(3) The S/N of the data should be at least $1.5$ at $\sim r_{500}$.  (4) The target should exhibit a relatively regular appearance and have no remarkable substructure to guarantee that the assumption of the spherical symmetry used in Section \ref{sec:model} is valid.
As a result we find 47 clusters and groups in the literature that satisfy the above conditions (Table \ref{sample}), the redshifts and averaged temperatures of which span the ranges $0.0036-0.63$ and $1.7-11.8$ keV, respectively.

\begin{startlongtable}
\begin{deluxetable}{llllllc}
	\tablecaption{The sample. \label{sample}}
	\tabletypesize{\footnotesize}
	\tablewidth{0pt}
	\tablehead{\colhead{Name} & \colhead{\Chandra{} Obsid\tablenotemark{a}} & \colhead{Detector} & \colhead{R.A.\tablenotemark{b}} & \colhead{Decl.\tablenotemark{b}}& \colhead{Redshift\tablenotemark{c}}   & \colhead{Reference\tablenotemark{d}} }
	\startdata
1E 1455.0+2232 &  4192 &  ACIS-I &  14:57:15.1 &  +22:20:34 & 0.258 & \cite{snowden08}\tablenotemark{2} \\
Abell 1068 &  1652 &  ACIS-S &  10:40:43.9 &  +39:56:53 & 0.147 &  \cite{snowden08}\tablenotemark{2}\\
Abell 1246 &  11770 &  ACIS-I &  11:23:50.0 &  +21:25:31 & 0.190 & \cite{sato14a}\tablenotemark{3} \\
Abell 133 &  9897 &  ACIS-I &  01:02:42.1 & $-$21:52:25 & 0.057 &  \cite{morandi14}\tablenotemark{1}\\
Abell 13 &  4945 &  ACIS-S &  00:13:38.3 & $-$19:30:08 & 0.103 &  \cite{snowden08}\tablenotemark{2}\\
Abell 1413 &  5003 &  ACIS-I &  11:55:18.9 &  +23:24:31 & 0.143 &  \cite{hoshino10}\tablenotemark{3}\\
Abell 1689 &  6930 &  ACIS-I &  13:11:29.5 & $-$01:20:17 & 0.183 &  \cite{kawa10}\tablenotemark{3}\\
Abell 1775 &  13510 &  ACIS-S &  13:41:53.8 &  +26:22:19 & 0.075 &  \cite{snowden08}\tablenotemark{2}\\
Abell 1795 &  10898 &  ACIS-I &  13:48:53.0 &  +26:35:44 & 0.062 & \cite{bautz09}\tablenotemark{3} \\
Abell 1835 &  6880 &  ACIS-I &  14:01:02.3 &  +02:52:48 & 0.253 & \cite{ichikawa13}\tablenotemark{3} \\
Abell 2029 &  4977 &  ACIS-S &  15:10:55.0 &  +05:43:12 & 0.077 &  \cite{walker12b}\tablenotemark{3}\\
Abell 209 &  3579 &  ACIS-I &  01:31:53.0 & $-$13:36:34 & 0.212 &  \cite{snowden08}\tablenotemark{2}\\
Abell 2142 &  5005 &  ACIS-I &  15:58:20.6 &  +27:13:37 & 0.091 & \cite{tcher16}\tablenotemark{2} \\
Abell 2163 &  1653 &  ACIS-I &  16:15:34.1 & $-$06:07:26 & 0.202 &  \cite{snowden08}\tablenotemark{2}\\
Abell 2199 &  10748 &  ACIS-I &  16:28:38.0 &  +39:32:55 & 0.030 &  \cite{sato14b}\tablenotemark{3}\\
Abell 2204 &  7940 &  ACIS-I &  16:32:46.5 &  +05:34:14 & 0.152 &  \cite{reiprich09}\tablenotemark{3}\\
Abell 2255 &  894 &  ACIS-I &  17:12:31.0 &  +64:05:33 & 0.081 &  \cite{akama17}\tablenotemark{3}\\
Abell 2319 &  3231 &  ACIS-I &  19:21:08.8 &  +43:57:30 & 0.056 & \cite{ghirar18}\tablenotemark{2} \\
Abell 2597 &  7329 &  ACIS-S &  23:25:20.0 & $-$12:07:38 & 0.080 & \cite{snowden08}\tablenotemark{2} \\
Abell 2667 &  2214 &  ACIS-S &  23:51:40.7 & $-$26:05:01 & 0.221 & \cite{snowden08}\tablenotemark{2} \\
Abell 3158 &  3712 &  ACIS-I &  03:42:53.9 & $-$53:38:07 & 0.060 &  \cite{ghirar18}\tablenotemark{2}\\
Abell 3266 &  899 &  ACIS-I &  04:31:24.1 & $-$61:26:38 & 0.059 &  \cite{ghirar18}\tablenotemark{2}\\
Abell 383 &  2320 &  ACIS-I &  02:48:02.0 & $-$03:32:15 & 0.187 &  \cite{snowden08}\tablenotemark{2}\\
Abell 478 &  1669 &  ACIS-S &  04:13:25.6 &  +10:28:01 & 0.088 &  \cite{point04}\tablenotemark{2}\\
Abell 644 &  2211 &  ACIS-I &  08:17:24.5 & $-$07:30:46 & 0.070 & \cite{ghirar18}\tablenotemark{2}\\
Abell 68 &  3250 &  ACIS-I &  00:37:05.3 &  +09:09:11 & 0.248 &  \cite{snowden08}\tablenotemark{2}\\
Abell 773 &  5006 &  ACIS-I &  09:17:59.4 &  +51:42:23 & 0.216 & \cite{snowden08}\tablenotemark{2} \\
Abell s1101 &  11758 &  ACIS-I &  23:13:58.6 & $-$42:44:02 & 0.056 &  \cite{snowden08}\tablenotemark{2}\\
Centaurus cluster &  4955 &  ACIS-S &  12:48:47.9 & $-$41:18:28 & 0.011 & \cite{walker13}\tablenotemark{3} \\
Cl 0016+16 &  520 &  ACIS-I &  00:18:33.8 &  +16:26:17 & 0.541 &  \cite{kotov05}\tablenotemark{2}\\
Cl 0024+17 &  929 &  ACIS-S &  00:26:35.7 &  +17:09:46 & 0.390 & \cite{kotov05}\tablenotemark{2} \\
Coma cluster &  13993 &  ACIS-I &  12:59:48.7 &  +27:58:50 & 0.023 &  \cite{simion13}\tablenotemark{3}\\
ESO 306- G 017 GROUP &  3188 &  ACIS-I &  05:40:06.3 & $-$40:50:32 & 0.036 & \cite{su13}\tablenotemark{3} \\
HydraA Cluster &  4970 &  ACIS-S &  09:18:06.5 & $-$12:05:36 & 0.054 & \cite{sato12}\tablenotemark{3} \\
Perseus cluster &  11714 &  ACIS-I &  03:19:47.2 &  +41:30:47 & 0.018 &  \cite{urban14}\tablenotemark{3}\\
Pks 0745-191 cluster &  6103 &  ACIS-I &  07:47:32.4 & $-$19:17:32 & 0.103 & \cite{walker12c}\tablenotemark{3} \\
Rxc j0605.8-3518 &  15315 &  ACIS-I &  06:05:52.8 & $-$35:18:02 & 0.137 & \cite{miller12}\tablenotemark{3} \\
Rxc j1825.3+3026 &  13381 &  ACIS-I &  18:25:24.7 &  +30:26:31 & 0.065 &  \cite{ghirar18}\tablenotemark{2}\\
Rxc j2234.5-3744 &  15303 &  ACIS-I &  22:34:31.0 & $-$37:44:06 & 0.154 & \cite{snowden08}\tablenotemark{2} \\
Rx j1120.1+4318 &  5771 &  ACIS-I &  11:20:07.4 &  +43:18:07 & 0.600 & \cite{kotov05}\tablenotemark{2} \\
Rx j1159.8+5531 &  4964 &  ACIS-S &  11:59:51.1 &  +55:31:56 & 0.081 &  \cite{su15}\tablenotemark{1}\\
Rx j1334.3+5030 &  5772 &  ACIS-I &  13:34:20.4 &  +50:31:05 & 0.620 & \cite{kotov05}\tablenotemark{2} \\
RX j1347.5-1145 &  3592 &  ACIS-I &  13:47:30.6 & $-$11:45:10 & 0.451 & \cite{snowden08}\tablenotemark{2}\\
Ugc 03957 cluster &  8265 &  ACIS-I &  07:40:58.3 &  +55:25:37 & 0.034 &  \cite{tholken16}\tablenotemark{3}\\
Virgo cluster &  7212 &  ACIS-I &  12:30:47.3 &  +12:20:13 & 0.0036 &  \cite{simion17}\tablenotemark{3}\\
Zwcl 1215.1+0400 &  4184 &  ACIS-I &  12:17:40.6 &  +03:39:45 & 0.075 & \cite{ghirar18}\tablenotemark{2}\\
Zwcl 3146 &  909 &  ACIS-I &  10:23:39.0 &  +04:11:14 & 0.282 &  \cite{snowden08}\tablenotemark{2}\\
\enddata
\tablenotetext{a}{The observation with the longest exposure time is chosen when multiple observations exist. Observations performed with ACIS-I are preferred if the target has been observed with both ACIS-I and ACIS-S.}
\tablenotetext{b}{Right ascensions and declinations are in J2000.0.}
\tablenotetext{c}{Redshifts are taken from NASA Extragalactic Database.}
\tablenotetext{d}{The superscripts 1, 2, and 3 in this column represent that the \Chandra{}, \XMM{}, and \Suzaku{} data are quoted from the literature, respectively.}
\end{deluxetable}
\end{startlongtable}

\section{{\it Chandra} Data Analysis}\label{sec:Chandra-data-analysis}
\subsection{Data Preparation}
The \Chandra{} data are reduced by following the method described in Z16. In brief, for each cluster or group we start from \Chandra{} ACIS level 1 event files and use the CIAO\footnote{Chandra Interactive Analysis of Observations, please refer to https://cxc.harvard.edu/ciao/.} script (version 4.11 with calibration database CALDB, v4.8.2) \texttt{chandra\_repro} to generate level 2 event files. After identifying and excluding point sources using the CIAO tool \texttt{celldetect}, the results of which have been cross-checked via visual examination, we examine the light curves extracted in $0.5-12.0$ keV from the source-free regions near the CCD edges, and filter the time intervals during which the count rate deviates from the mean value by $20\%$.
\subsection{Data Analysis}
In this subsection we calculate the X-ray surface brightness and gas temperature profiles to be used in the RTI model fitting (Section \ref{sec:model}) by fitting the \Chandra{} ACIS data.  
We first extract the surface brightness profiles in $0.7-7.0$ keV from a set of concentric annuli that are centered at the X-ray centroid and use the exposure maps generated with the CIAO tool \texttt{flux\_image} to correct for the vignetting effect and exposure fluctuations.   
Then, we derive the gas temperature profiles by analyzing the spectra extracted in $0.7-7$ keV from another set of annuli, which are also centered at the X-ray centroid but are wider to include a minimal of 2500 photon counts per annulus. The extracted spectra are corrected using the Redistribution Matrix Files (RMFs) and Ancillary Response Files (ARFs) generated with the CIAO script \texttt{specextract} after subtracting the backgrounds, which are created by following the method given in Z16. To fit the spectra we apply the multiplication of XSPEC models (1) \texttt{PROJCT} to correct for the deprojection effect, (2) two \texttt{APEC} components (one for the thermal emission of the optically thin ICM, and the other for the possible cool phase gas that is often observed in the central region; e.g., \citealt{makishima01}. The abundances of both \texttt{APEC} components are set free unless either or both of them cannot be well constrained; in this case we fix the abundance of the corresponding component to 0.3 solar; e.g., \citealt{panagoulia14}), and (3) \texttt{WABS} to model the photoelectric absorption, which is fixed to the Galactic value \citep{kalberla05}. If $F$-test shows that the two-phase fitting is not significantly better ($p$-value > 0.05) than the case when the cool phase is ignored, we turn to choose the single-phase gas model.

\section{RTI model analysis of the sample} \label{sec:model}
\subsection{RTI Model Description}
In this section we introduce the RTI model used to fit the observed gas temperature and surface brightness profiles, which is an improved version of the model we presented in Z16. Similar to what we have done in Z16, for a given test element that contains $n^*$ gas particles we assume that the current total energy contained in the gas is supplied by (1) the energy injected through gravitational collapse ($\Delta E_{\rm G}(r)$), (2) the net heat obtained in its thermal history ($\Delta E_{\rm heating}(r)$) via feedback processes, radiative cooling, thermal conduction, etc., (3) the work done by surrounding particles through volume change ($\Delta E_{\rm work}(r)$). Since the current gas thermal energy occupies only part of $\Delta E_{\rm G}(r)+\Delta E_{\rm heating}(r)+\Delta E_{\rm work}(r)$ (the rest is stored in turbulence or bulk motions), it can be found that (c.f., equations 6 and 13 in Z16) the gas temperature profile should have the form   
\begin{equation}\label{T_gen}
T(r)=\frac{2\eta(r)(E_0+\Delta E_{\rm G}(r)+\Delta E_{\rm work}(r)+\Delta E_{\rm heating}(r))}{3n^*k_{\rm B}},
\end{equation}
where $\eta(r)$ is the fraction of the thermal pressure to the total pressure, $E_0$ ($\sim 0$) represents the initial energy of the gas element, and $k_{\rm B}$ is the Boltzmann constant.

\noindent\textbf{Energy injected through gravitational collapse:} To calculate $\Delta E_{\rm G}(r)$ we assume a generalized NFW profile \citep{nfw97} to describe the total gravitational mass distribution, which is 
\begin{equation}\label{MNFW}
\rho(r)=\frac{\rho_0}{(r/r_s)^{\delta_1}(1+r/r_s)^{\delta_2-\delta_1}},
\end{equation}
where $r_s$ is the scale radius, $\rho_0$ is the central density, and $\delta_1$/$\delta_2$ represents the inner/outer slope. 
Thus the corresponding total energy input via gravitational collapse (c.f. Equation 7 in Z16) can be calculated as 
\begin{equation}\label{E_G}
\Delta E_{\rm G}(r) = \frac{G\int_{0}^{r}\rho(x)4\pi x^2m^*dx}{r}+\int_{r}^{\infty}\frac{G\rho(x)4\pi x^2m^*dx}{x},
\end{equation}
where $G$ is the gravitational constant, $m^*=n^*\mu m_{\rm p}$ is the mass of the test gas element ($m_{\rm p}$ is the proton mass and $\mu =0.61$ is the mean molecular weight).

\noindent\textbf{Net heat supply:} The net heat supply $\Delta E_{\rm heating}(r)$ is estimated using the thermodynamic relation $dQ=TdS$, where $S$ is the thermodynamical entropy. Here we adopt the Boltzmann entropy for the ideal gas, one of the most frequently used thermodynamical entropies (they differ from each other by only a constant), to calculate $dQ$ \citep{kardar07}, which is
\begin{equation}
S=n^*k_{\rm B}(1.5+1.5\ln(2\pi m^*k_{\rm B}T)-\ln(n^*/V^*)),
\end{equation}
where $V^*$ is the volume of the test gas element. The change of entropy between any states 1 and 2 is then written as
\begin{equation}\label{Eq:S-diff}
S_2-S_1=n^*k_{\rm B}\ln(\frac{T_2^{1.5}V^*_2}{T_1^{1.5}V^*_1}),
\end{equation}
or, considering that the conventional definition of astrophysical entropy is $K=Tn_{\rm e}^{-2/3}$, 
\begin{equation}
S_2-S_1=1.5n^*k_{\rm B}\ln(K_2/K_1).
\end{equation} 
Taking the limit state 2 $\rightarrow$ state 1 we have $dS=1.5n^*k_{\rm B}dK/K$. Therefore, by substituting $dS$ into the expression $dQ=TdS$ the net heat supply in the non-gravitational processes becomes 
\begin{equation}\label{Eq:E_H-def}
\Delta E_{\rm heating}(r)=\int dQ(r)=\int_{K_{\rm sim}}^{K_{\rm obs}}1.5n^*k_{\rm B}T(r)dK/K(r),
\end{equation}
where $K_{\rm sim}$ is the simulated entropy of the gas element if only the gravitational effect is considered \citep{voit05b}, and  $K_{\rm obs}$ is the observed entropy. We estimate the integration in Equation \ref{Eq:E_H-def} using the method of \citet{cnm12},   
\begin{equation}\label{E_H}
\Delta E_{\rm heating}(r) = C_hn^*k_{\rm B}T(r) (K_{\rm obs}(r) - K_{\rm sim}(r)) / K_{\rm obs}(r), 
\end{equation} 
where $C_h$ is the scaling factor. \citet{voit05b} showed that when only the gravitational effect is considered (see their equation 5), 
\begin{equation}\label{kmodel}
K_{\rm sim}(r)=K_0+A_0\times \frac{r}{\rm 1\ kpc}^{\gamma_0},
\end{equation}
where $K_0$ is the central entropy, $A_0$ is the normalization and $\gamma_0$ is the slope of the entropy profile. To express the observed entropy $K_{\rm obs}(r)$, which will be derived later by fitting the observed gas temperature and X-ray surface brightness with the model described here, we adopt the electron density profile $n_{\rm e}(r)$ following \citet{patej15} 
\begin{equation}\label{ne}
n_{\rm e}(r)=\frac{\Gamma f_g}{\mathbf{\mu_{\rm e}m_{\rm p}}}\left(\frac{r}{r_{\rm shock}}\right)^{3\Gamma-3}\rho(r_{\rm shock}\left[\frac{r}{r_{\rm shock}}\right]^{\Gamma}),
\end{equation}
where $r_{\rm shock}$ is the radius where virial shock happens, $\mu_{\rm e}=1.18$ is the mean molecular weight per electron, $\Gamma$ is the ratio used to measure the gas density jump at $r_{\rm shock}$, $f_g$ is the gas fraction within $r_{\rm shock}$, and $\rho$ is the total mass density (Equation \ref{MNFW}). 

\noindent\textbf{Work done by surrounding particles:} We calculate $\Delta E_{\rm work}$ by assuming that the gas element has experienced a polytropic process during the gravitational collapse. Thus, we calculate $\Delta E_{\rm work}(r)$ (c.f., equation 9 in Z16) as
\begin{equation}\label{E_W}
\Delta E_{\rm work}(r)=\frac{\nu^*_{\rm mol} \mathcal{R}}{n-1}(T(r)-\frac{2C_w\Delta E_{\rm G}(r)}{3n^*k_{\rm B}}),
\end{equation}
where $n$ is the polytropic index and $C_w$ is the fraction of the gravitational energy that has been transformed into the thermal energy before the polytropic process, both of which are to be determined in the fitting, and $\nu_{\rm mol}^*=n^*/N_A$ is the mole number ($N_A$ is the Avogadro constant).

\noindent\textbf{Model predicted gas temperature and X-ray surface brightness:} Following \citet{nel14} we write the thermal pressure fraction $\eta(r)$ in Equation \ref{T_gen} as
\begin{equation}\label{eta}
\eta(r)=\frac{P_{\rm thermal}}{P_{\rm tot}}=A\left(1+\exp\left[-\left(\frac{r/r_{200}}{B}\right)^\gamma\right]\right),
\end{equation}
where $P_{\rm thermal}$ and $P_{\rm tot}$ are the thermal and total pressure, respectively ($A$, $B$, and $\gamma$ are empirical parameters calculated in \citet{nel14}). Note that when $\eta(r)=1$ (cases 3, 4 and 5 in Section \ref{sec:model-comp}) the non-thermal pressure vanished. Now, the model-predicted gas temperature distribution can be obtained by substituting Equations \ref{E_G}, \ref{E_H}, \ref{E_W}, and \ref{eta} into Equation \ref{T_gen}. To write it in a simple analytical form, we introduce $e_{\rm gen}(r)\equiv \Delta E_{\rm G}(r)/\Delta E_{\rm G,ref}(r)$, where the reference term $\Delta E_{\rm G,ref}(r)$ is calculated using Equation \ref{E_G} by setting $\delta_1 \equiv 1$ and $\delta_2 \equiv 3$ (when $\delta_1$ and $\delta_2$ are not integers, $\Delta E_{\rm G}(r)$ has no analytical expression). Thus,
\begin{equation}\label{Tfin}
T(r)=\frac{T_0+e_{\rm gen}(r)(1-C_wN_3)N_1\rho_0r_s^3\ln((r_s+r)/r_s)/r-K_{\rm sim}(r)/n_{\rm e}^{-2/3}(r)}{1/\eta(r)-N_3-N_2},
\end{equation}
where 
$T_{0}=2E_0/3k_{\rm B}$ is assumed to be $\simeq 0$, $N_1=8\pi G\mu m_{\rm p}/3$ is a fixed combination of physical constants, $N_2=2C_h/3$, and $N_3=2/(3(n-1))$. 
The 3-dimensional temperature profiles calculated using Equation \ref{Tfin} or the corresponding 2-dimensional profiles obtained by projecting the 3-dimensional profiles using the method of \citet{mazza04}, i.e.,
\begin{equation}\label{Eq:T2d}
T_{\rm 2d}(R)=\frac{\int_{R}^{\infty} n_{\rm e}^2(r) T^{0.25}(r)dr}{\int_{R}^{\infty} n_{\rm e}^2(r)T^{-0.75}(r)dr},
\end{equation}
can be directly compared with the observed temperatures. 

The model-predicted X-ray surface brightness profile is
\begin{equation}\label{sbp}
S_{\rm model}(R)=\int_{R}^{\infty}C(r)\Lambda(T,Z)\bar{n}_{\rm e}\bar{n}_{\rm p}(r)\frac{2rdr}{\sqrt{r^2-R^2}}+S_{\rm bkg},
\end{equation}
where $n_{\rm p}$ is the proton number density ($\simeq n_{\rm e}/1.2$ for a fully ionized ICM; \citealt{cava09}), $\Lambda(T,Z)$ is the cooling function calculated using the temperature derived from Equation \ref{Tfin} and the abundance given in the references listed in Table \ref{sample}, $S_{\rm bkg}$ is the diffuse X-ray background, and $C(r)$ is the clumping factor to characterize the level of inhomogeneities in the ICM. In literature $C(r)$ is usually defined as 
\begin{equation}\label{C_def}
C(r)\equiv \frac{\langle\rho_{\rm gas}^2(r)\rangle}{\langle\rho_{\rm gas}(r)\rangle^2},
\end{equation} 
where $\langle\ \cdot \  \rangle$ denotes the average operation inside a spherical shell, so that $C(r)=1$ (cases 2 to 5 in Section \ref{sec:model-comp}) when the clumping effect does not exist.
{In the high-resolution cosmological simulations of \citet{vazza13}, where the effects of AGN feedback and gas cooling were both considered, the radial variation of the clumping factor was found to obey the following empirical relation,}
\begin{equation}\label{C_prof}
C(r)^{0.5}=(1+r/r_{200})^{c_1}\exp({c_2r/r_{200}})+c_3\exp\left[\frac{-(r/r_{200}-c_4)^2}{c_5}\right],
\end{equation}  
where $c_1$ to $c_5$ are free parameters to be determined in the model fitting. 


\subsection{Markov-chain Monte Carlo Sampling and Results of RTI Model Fitting}\label{sec:mcmc}
\begin{deluxetable}{lccccc}
	\tablecaption{Priors of model parameters.\label{tab:prior}}
	\tablehead{\colhead{Parameter} & \colhead{Mean\tablenotemark{a}} & \colhead{Standard deviation\tablenotemark{a}} & \colhead{Min} & \colhead{Max} & \colhead{Reference}}
	\startdata
	$\rho_0$ ($M_\odot$ ${\rm kpc^{-3}}$) & N/A & N/A & $0$ &  $10^7$ & N/A \\
	$r_s$ (Mpc)& N/A & N/A & $0$ & $1$ & N/A \\
	$\delta_1$& $1.0$ & $0.5$ & $0$ & $2$ & \cite{nfw97} \\
	$\delta_2$& $3.0$ & $0.5$ & $2$ & $10$ & \cite{nfw97} \\
	$A_0$ (keV $\rm cm^{2}$) & $0.5$ & $1.0$ & $0$ & $5$ & \cite{voit05b} \\
	$\gamma_0$ & $1.1$ & $0.2$ & $0$ & $3$ & \cite{voit05b} \\
	$K_0$ (keV $\rm cm^{2}$)& N/A & N/A & $0$ & $1000$ & \cite{voit05b} \\
	$\Gamma$ & $1.3$ & $0.5$ & $1.0$ & $5.0$ & \cite{patej15} \\
	$f_g$ & $0.13$ & $0.03$ & $0$ & $0.3$ & \cite{patej15} \\
	$r_{\rm shock}$ (Mpc) & N/A & N/A & $2$ & $10$ & \cite{patej15} \\  
	$A$ & $0.45$ & $0.2$ & $0$ & $0.5$ & \cite{nel14} \\
	$B$ & $1.4$ & $0.7$ & $0$ & N/A & \cite{nel14} \\
	$\gamma$ & $1.6$ & $0.8$ & $0$ & N/A& \cite{nel14} \\
	$N_2$ & $1.0$ & $0.2$ & $0$ & $5$ & \cite{cnm12} \\
	$N_3$ & N/A & N/A& $0$ & $10$ & N/A \\
	$C_w$ & N/A & N/A & $0$ & $1$ & N/A \\
	$c_1$ & $-3.7$ & $1.0$ & $-10$ & $10$ & \cite{vazza13}\\ 
	$c_2$ & $3.7$ & $1.0$ & $-10$ & $10$ & \cite{vazza13}\\ 
	$c_3$ & N/A & N/A & $0$ & $5$ & \cite{vazza13}\\ 
	$c_4$ & N/A & N/A & $0$ & $0.1$ & \cite{vazza13}\\ 
	$c_5$ & N/A & N/A & $0$ & $0.1$ & \cite{vazza13}\\ 
	\enddata
	\tablenotetext{a}{When the mean and standard deviation for a model parameter is not available, we adopt a Uniform prior (i.e., no prior is assumed). Otherwise a Gaussian distribution is assigned (e.g., \citealp{andreon13}).}
\end{deluxetable}
Similar to Z16 we employ the Bayesian approach (see \citealt{andreon13} for a review), which can be used to quantify the intrinsic scatters and uncertainties of known model parameters (i.e., all parameters listed in Table \ref{tab:prior}) 
meanwhile incorporate observation errors (i.e., both statistic and systematic errors of observed temperature and surface brightness; a more detailed discussion on systematic errors is presented in Section \ref{sec:err}) in the model fittings. In the Bayesian approach, Markov-chain Monte Carlo (MCMC) algorithm is often adopted since it is possible to use it to perform straightforward fittings to a  complicated model with huge numerical calculations at a moderate convergence speed. After the convergence is achieved, the MCMC algorithm iteratively generates a large number of samples from the true joint distribution of the model parameters \citep{trotta17} that can be used to estimate the possibility distributions of all model parameters (functions of these parameters). In this work we employ the Metropolis–Hastings MCMC \citep{Bimka70} that is implemented by the software PyMC\footnote{https://github.com/pymc-devs/pymc} to obtain the best-fit model parameters by maximizing the posterior (i.e., the possibility of the fitted parameters for given observed data sets), which is calculated by multiplying the prior (i.e., our knowledge of the model parameters obtained before the fittings are carried out; see Table \ref{tab:prior}) and the likelihood function $L$ (i.e., the possibility of the observed data; c.f., \citealt{andreon13}) 
\begin{equation}\label{eq:lhood}
L=\prod_{i} \frac{1}{\sqrt{2\pi \sigma_{x,i}^2}}\exp\left[-\frac{\left(x_{i,{\rm obs}}-x_{i,{\rm model}}\right)^2}{2\sigma_{x,i}^2}\right],
\end{equation}
where $x_{i,{\rm obs}}$ represents the observed variables including gas temperature and X-ray surface brightness, as well as the total gravitating mass, 
$\sigma_{x,i}$ represents the standard deviation of the corresponding $x_{i,{\rm obs}}$, and $x_{i,\rm model}$ represents the corresponding model-predicted value. 
We have performed a split test \citep{morandi16} to ensure the convergence of the MCMC chains, i.e., for each sample member the MCMC chain is split into two series after the burn-in (i.e., the first part of iterations in the MCMC chain before it converges) is removed, and the model parameters derived with the two split chains are compared with each other to make sure their $68\%$ confidence range overlap each other. The number of MCMC iterations is set to $5 \times 10^5$, and the first $ 2\times 10^5$ iterations are regarded as burn-in to ensure the convergence.

\begin{startlongtable}
	\begin{deluxetable}{lcccccc}
		\tablecaption{Best-fit results.\label{sample_result}}
		\tabletypesize{\scriptsize}
		\tablewidth{0pt}
 	 \tablehead{\colhead{Name} & \colhead{$r_{200}$ } & \colhead{Averaged temperature\tablenotemark{a}} &\colhead{$M_{200}$} &\colhead{$M_{\rm gas,200}$} &\colhead{ $L_{X,200}$\tablenotemark{b}} & {$ R_{\rm eff}$\tablenotemark{c}}\\
	&(Mpc) & (keV) &($10^{14}$ $M_\odot$) & ($10^{13}$ $M_\odot$) & \quad \;\,($10^{44}$ erg $\rm s^{-1}$) &   }
		\startdata
 1E 1455.0+2232 & $1.52_{-0.09}^{+0.10}$ & $5.23_{-0.27}^{+0.26}$ & $5.16_{-0.85}^{+1.14}$ & $6.65_{-0.65}^{+0.70}$ & $16.74_{-0.46}^{+0.46}$ & $0.38$ \\
Abell 1068 & $1.53_{-0.06}^{+0.06}$ & $4.48_{-0.16}^{+0.16}$ & $4.79_{-0.56}^{+0.54}$ & $5.41_{-0.47}^{+0.47}$ & $8.07_{-0.18}^{+0.25}$ & $0.55$ \\
Abell 1246 & $1.62_{-0.07}^{+0.08}$ & $6.37_{-0.37}^{+0.43}$ & $5.95_{-0.76}^{+0.92}$ & $7.61_{-1.02}^{+0.90}$ & $6.60_{-0.35}^{+0.32}$ & $0.33$ \\
Abell 13 & $1.51_{-0.06}^{+0.06}$ & $4.66_{-0.18}^{+0.17}$ & $4.39_{-0.49}^{+0.56}$ & $5.92_{-0.52}^{+0.50}$ & $2.52_{-0.13}^{+0.13}$ & $0.52$ \\
Abell 133 & $1.47_{-0.06}^{+0.06}$ & $3.95_{-0.12}^{+0.15}$ & $3.89_{-0.47}^{+0.49}$ & $4.09_{-0.31}^{+0.32}$ & $3.14_{-0.13}^{+0.13}$ & $0.45$ \\
Abell 1413 & $1.82_{-0.07}^{+0.08}$ & $7.01_{-0.35}^{+0.31}$ & $7.94_{-0.85}^{+1.07}$ & $9.50_{-0.88}^{+0.75}$ & $13.01_{-0.42}^{+0.34}$ & $0.62$ \\
Abell 1689 & $1.99_{-0.07}^{+0.09}$ & $9.16_{-0.39}^{+0.41}$ & $10.79_{-1.13}^{+1.49}$ & $14.92_{-0.93}^{+0.85}$ & $28.46_{-0.75}^{+0.73}$ & $0.52$ \\
Abell 1775 & $1.41_{-0.05}^{+0.06}$ & $3.89_{-0.13}^{+0.12}$ & $3.47_{-0.38}^{+0.47}$ & $4.05_{-0.46}^{+0.42}$ & $2.36_{-0.11}^{+0.11}$ & $0.54$ \\
Abell 1795 & $1.70_{-0.05}^{+0.06}$ & $5.25_{-0.18}^{+0.18}$ & $6.06_{-0.55}^{+0.62}$ & $8.71_{-0.72}^{+0.81}$ & $12.04_{-0.38}^{+0.37}$ & $0.66$ \\
Abell 1835 & $1.82_{-0.05}^{+0.07}$ & $7.41_{-0.34}^{+0.35}$ & $8.93_{-0.64}^{+1.14}$ & $13.26_{-1.00}^{+1.35}$ & $42.85_{-1.23}^{+1.39}$ & $0.45$ \\
Abell 2029 & $2.17_{-0.09}^{+0.08}$ & $5.90_{-0.20}^{+0.20}$ & $12.77_{-1.47}^{+1.47}$ & $13.90_{-1.01}^{+1.19}$ & $20.28_{-0.63}^{+0.69}$ & $0.01$ \\
Abell 209 & $1.80_{-0.07}^{+0.09}$ & $6.98_{-0.31}^{+0.32}$ & $8.29_{-0.98}^{+1.38}$ & $12.18_{-1.40}^{+1.84}$ & $13.60_{-0.67}^{+0.60}$ & $0.64$ \\
Abell 2142 & $2.18_{-0.05}^{+0.06}$ & $8.03_{-0.14}^{+0.14}$ & $13.07_{-0.93}^{+1.03}$ & $20.66_{-1.22}^{+1.63}$ & $26.70_{-0.68}^{+0.67}$ & $0.72$ \\
Abell 2163 & $2.88_{-0.15}^{+0.11}$ & $12.08_{-0.50}^{+0.41}$ & $33.39_{-4.91}^{+4.13}$ & $50.48_{-6.54}^{+4.73}$ & $60.15_{-4.11}^{+2.55}$ & $0.40$ \\
Abell 2199 & $1.38_{-0.05}^{+0.06}$ & $3.96_{-0.14}^{+0.15}$ & $3.14_{-0.33}^{+0.40}$ & $3.21_{-0.23}^{+0.28}$ & $3.53_{-0.10}^{+0.13}$ & $0.69$ \\
Abell 2204 & $1.91_{-0.07}^{+0.07}$ & $7.18_{-0.33}^{+0.29}$ & $9.27_{-0.93}^{+1.11}$ & $13.15_{-1.27}^{+1.39}$ & $32.06_{-1.06}^{+1.04}$ & $0.62$ \\
Abell 2255 & $1.51_{-0.09}^{+0.10}$ & $5.77_{-0.30}^{+0.30}$ & $4.34_{-0.71}^{+0.89}$ & $4.33_{-0.91}^{+1.31}$ & $4.51_{-0.30}^{+0.29}$ & $0.56$ \\
Abell 2319 & $2.17_{-0.07}^{+0.08}$ & $8.28_{-0.24}^{+0.20}$ & $12.51_{-1.13}^{+1.34}$ & $19.99_{-1.67}^{+2.00}$ & $18.37_{-0.61}^{+0.72}$ & $0.76$ \\
Abell 2597 & $1.36_{-0.05}^{+0.05}$ & $3.56_{-0.12}^{+0.11}$ & $3.14_{-0.33}^{+0.38}$ & $3.50_{-0.24}^{+0.24}$ & $5.43_{-0.12}^{+0.14}$ & $0.55$ \\
Abell 2667 & $1.80_{-0.07}^{+0.08}$ & $6.89_{-0.31}^{+0.31}$ & $8.34_{-0.98}^{+1.15}$ & $10.58_{-1.27}^{+1.16}$ & $23.27_{-0.82}^{+0.72}$ & $0.50$ \\
Abell 3158 & $1.56_{-0.07}^{+0.07}$ & $4.89_{-0.14}^{+0.15}$ & $4.67_{-0.57}^{+0.69}$ & $5.12_{-0.63}^{+0.71}$ & $4.63_{-0.22}^{+0.22}$ & $0.67$ \\
Abell 3266 & $1.53_{-0.21}^{+0.14}$ & $6.34_{-0.47}^{+0.46}$ & $4.42_{-1.56}^{+1.34}$ & $6.57_{-2.74}^{+2.60}$ & $7.45_{-0.53}^{+0.55}$ & $0.33$ \\
Abell 383 & $1.55_{-0.08}^{+0.08}$ & $4.78_{-0.18}^{+0.21}$ & $5.19_{-0.73}^{+0.80}$ & $5.76_{-0.43}^{+0.41}$ & $8.29_{-0.23}^{+0.24}$ & $0.77$ \\
Abell 478 & $1.87_{-0.09}^{+0.09}$ & $6.29_{-0.30}^{+0.25}$ & $8.25_{-1.12}^{+1.21}$ & $10.10_{-1.00}^{+1.11}$ & $25.62_{-0.60}^{+0.58}$ & $0.44$ \\
Abell 644 & $1.82_{-0.05}^{+0.05}$ & $6.51_{-0.22}^{+0.20}$ & $7.42_{-0.62}^{+0.62}$ & $11.33_{-0.71}^{+0.61}$ & $11.12_{-0.34}^{+0.34}$ & $0.63$ \\
Abell 68 & $1.68_{-0.09}^{+0.08}$ & $6.72_{-0.38}^{+0.36}$ & $7.03_{-1.06}^{+1.05}$ & $9.72_{-1.98}^{+1.36}$ & $11.94_{-0.60}^{+0.54}$ & $0.47$ \\
Abell 773 & $1.74_{-0.08}^{+0.09}$ & $7.27_{-0.31}^{+0.34}$ & $7.42_{-0.97}^{+1.18}$ & $9.48_{-1.18}^{+1.29}$ & $12.93_{-0.56}^{+0.47}$ & $0.55$ \\
Abell s1101 & $1.13_{-0.04}^{+0.05}$ & $2.44_{-0.08}^{+0.08}$ & $1.76_{-0.18}^{+0.23}$ & $2.17_{-0.11}^{+0.12}$ & $2.35_{-0.05}^{+0.05}$ & $0.69$ \\
Centaurus cluster & $1.25_{-0.04}^{+0.06}$ & $2.91_{-0.11}^{+0.11}$ & $2.33_{-0.24}^{+0.33}$ & $2.41_{-0.21}^{+0.23}$ & $1.05_{-0.07}^{+0.04}$ & $0.28$ \\
Cl 0016+16 & $1.74_{-0.07}^{+0.09}$ & $9.36_{-0.51}^{+0.52}$ & $10.68_{-1.27}^{+1.74}$ & $15.38_{-2.03}^{+2.89}$ & $36.76_{-1.24}^{+1.66}$ & $0.46$ \\
Cl 0024+17 & $1.14_{-0.05}^{+0.05}$ & $3.41_{-0.20}^{+0.20}$ & $2.52_{-0.30}^{+0.37}$ & $3.37_{-0.53}^{+0.45}$ & $2.85_{-0.23}^{+0.23}$ & $0.52$ \\
Coma cluster & $1.96_{-0.10}^{+0.09}$ & $7.76_{-0.23}^{+0.25}$ & $9.00_{-1.26}^{+1.36}$ & $10.47_{-1.09}^{+1.38}$ & $9.07_{-0.31}^{+0.32}$ & $0.55$ \\
ESO 306- G 017 GROUP & $1.14_{-0.06}^{+0.05}$ & $2.43_{-0.12}^{+0.12}$ & $1.77_{-0.27}^{+0.24}$ & $1.60_{-0.16}^{+0.17}$ & $0.59_{-0.04}^{+0.04}$ & $0.47$ \\
HydraA Cluster & $1.32_{-0.04}^{+0.04}$ & $3.27_{-0.15}^{+0.13}$ & $2.84_{-0.27}^{+0.30}$ & $4.05_{-0.30}^{+0.30}$ & $5.12_{-0.14}^{+0.13}$ & $0.68$ \\
Perseus cluster & $1.83_{-0.07}^{+0.06}$ & $6.12_{-0.15}^{+0.14}$ & $7.26_{-0.78}^{+0.81}$ & $9.37_{-0.60}^{+0.57}$ & $12.48_{-0.21}^{+0.25}$ & $0.60$ \\
Pks 0745-191 cluster & $2.10_{-0.10}^{+0.12}$ & $8.15_{-0.37}^{+0.35}$ & $11.89_{-1.59}^{+2.21}$ & $14.99_{-1.47}^{+1.46}$ & $30.73_{-0.87}^{+0.93}$ & $0.71$ \\
Rxc j0605.8-3518 & $1.60_{-0.08}^{+0.08}$ & $4.87_{-0.21}^{+0.19}$ & $5.36_{-0.72}^{+0.82}$ & $6.28_{-0.64}^{+0.65}$ & $8.14_{-0.33}^{+0.23}$ & $0.71$ \\
Rxc j1825.3+3026 & $1.52_{-0.08}^{+0.08}$ & $5.58_{-0.23}^{+0.21}$ & $4.30_{-0.64}^{+0.74}$ & $4.54_{-0.97}^{+1.03}$ & $3.07_{-0.26}^{+0.25}$ & $0.57$ \\
Rxc j2234.5-3744 & $1.76_{-0.06}^{+0.07}$ & $7.75_{-0.38}^{+0.36}$ & $7.24_{-0.69}^{+0.93}$ & $10.34_{-1.26}^{+0.89}$ & $13.11_{-0.42}^{+0.36}$ & $0.63$ \\
Rx j1120.1+4318 & $1.10_{-0.08}^{+0.07}$ & $4.70_{-0.36}^{+0.38}$ & $2.83_{-0.60}^{+0.62}$ & $2.96_{-0.69}^{+0.78}$ & $10.50_{-0.67}^{+0.55}$ & $0.50$ \\
Rx j1159.8+5531 & $0.93_{-0.03}^{+0.04}$ & $1.67_{-0.06}^{+0.06}$ & $1.00_{-0.10}^{+0.12}$ & $0.80_{-0.05}^{+0.07}$ & $0.24_{-0.01}^{+0.01}$ & $0.27$ \\
Rx j1334.3+5030 & $1.16_{-0.08}^{+0.08}$ & $4.84_{-0.36}^{+0.41}$ & $3.44_{-0.70}^{+0.77}$ & $4.22_{-1.32}^{+1.35}$ & $7.10_{-0.74}^{+0.63}$ & $0.50$ \\
RX j1347.5-1145 & $2.11_{-0.10}^{+0.11}$ & $12.19_{-0.53}^{+0.73}$ & $17.14_{-2.21}^{+2.95}$ & $23.61_{-2.15}^{+2.22}$ & $103.48_{-3.16}^{+3.26}$ & $0.36$ \\
Ugc 03957 cluster & $1.06_{-0.03}^{+0.03}$ & $2.23_{-0.08}^{+0.08}$ & $1.45_{-0.13}^{+0.12}$ & $1.12_{-0.09}^{+0.08}$ & $0.69_{-0.03}^{+0.03}$ & $0.61$ \\
Virgo cluster & $1.08_{-0.03}^{+0.03}$ & $2.21_{-0.05}^{+0.06}$ & $1.48_{-0.12}^{+0.12}$ & $1.03_{-0.08}^{+0.09}$ & $0.42_{-0.02}^{+0.02}$ & $0.52$ \\
Zwcl 1215.1+0400 & $1.74_{-0.06}^{+0.07}$ & $6.15_{-0.21}^{+0.20}$ & $6.51_{-0.69}^{+0.79}$ & $8.39_{-0.77}^{+0.61}$ & $5.26_{-0.23}^{+0.27}$ & $0.66$ \\
Zwcl 3146 & $1.90_{-0.08}^{+0.10}$ & $7.49_{-0.31}^{+0.40}$ & $10.35_{-1.21}^{+1.77}$ & $13.56_{-0.92}^{+0.90}$ & $34.68_{-0.93}^{+0.76}$ & $0.69$ \\
		\enddata
		\tablenotetext{a}{The volume-averaged temperature is calculated for the $0.2-0.5$ $r_{500}$ region using the best-fit temperature profile.}
		\tablenotetext{b}{$L_{\rm X,200}$ is the X-ray luminosity within $r_{200}$ calculated in the $0.1-50$ keV.}
		\tablenotetext{c}{$R_{\rm eff}$ is used to {describe} the goodness of fitting (Equation \ref{eff}).}
	\end{deluxetable}
\end{startlongtable}

In Table \ref{sample_result} we list best-fit results together with the goodness of fit described with the model efficiency $R_{\rm eff}$ (see Z16 and references therein)
\begin{equation}\label{eff}
R_{\rm eff}=\frac{1}{N_{\rm bin}}\sum_{i=1}^{N_{\rm bin}}R_{{\rm eff},i}, 
\end{equation}
in which
\begin{equation}
R_{\rm eff,i}=1-\frac{\sum_{n=1}^{N_{\rm sample}}\left(X_{i,n,\rm sample}-\bar{X}_{i,\rm obs}\right)^2}{\sum_{n=1}^{N_{\rm sample}}\left(X_{i,n,\rm obs}-\bar{X}_{i,\rm obs}\right)^2},
\end{equation}   
where $N_{\rm bin}$ is the total bin number of the observed quantities (i.e., temperature, surface brightness, and total mass), $N_{\rm sample}$ is the number of MCMC iterations, and $X_{\rm sample}$ and $X_{\rm obs}$ are used to represent any sampled and observed physical quantities, respectively.
\begin{figure}
	\includegraphics[width=1.0\textwidth]{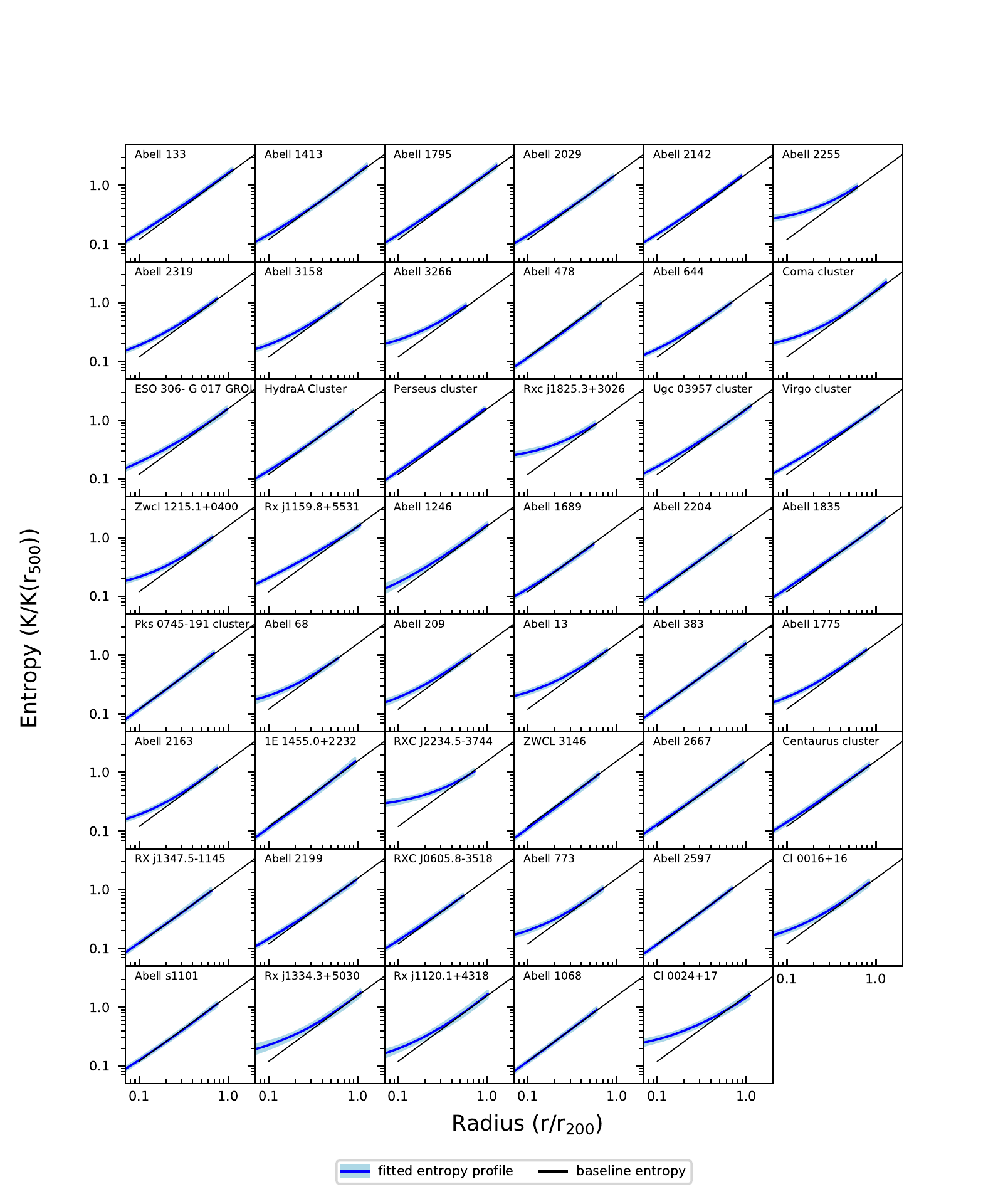}
	\centering
	\caption{Gas entropy profiles derived from the converged MCMC chains (blue lines) together with the $68\%$ error range (shaded regions). The simulation-predicted profiles (black line) are also plotted for comparison. All the profiles have been scaled by $r_{200}$ and $K(r_{500})$). \label{fig:sample} }
\end{figure}
We find that a reasonable fit ($R_{\rm eff}\gtrsim 0$; if the fitting is based on the $\chi$-square statistic, $R_{\rm eff}\gtrsim 0$ implies $\chi^2_{\nu} \lesssim 1$; see also \citealt{engeland02}) has been obtained for all targets in the sample. Then we calculate the entropy profiles using the best-fit gas densities and temperature profiles, and plot them in Figure \ref{fig:sample}, with the simulation-predicted entropy profiles \citep{voit05b}, which are scaled by $r_{200}$ and $K(r_{500})$. It showed that no apparent flattening of the gas entropy profile can be confirmed near $\sim r_{200}$. 

\begin{figure}
	\centering
	\includegraphics[width=0.8\textwidth]{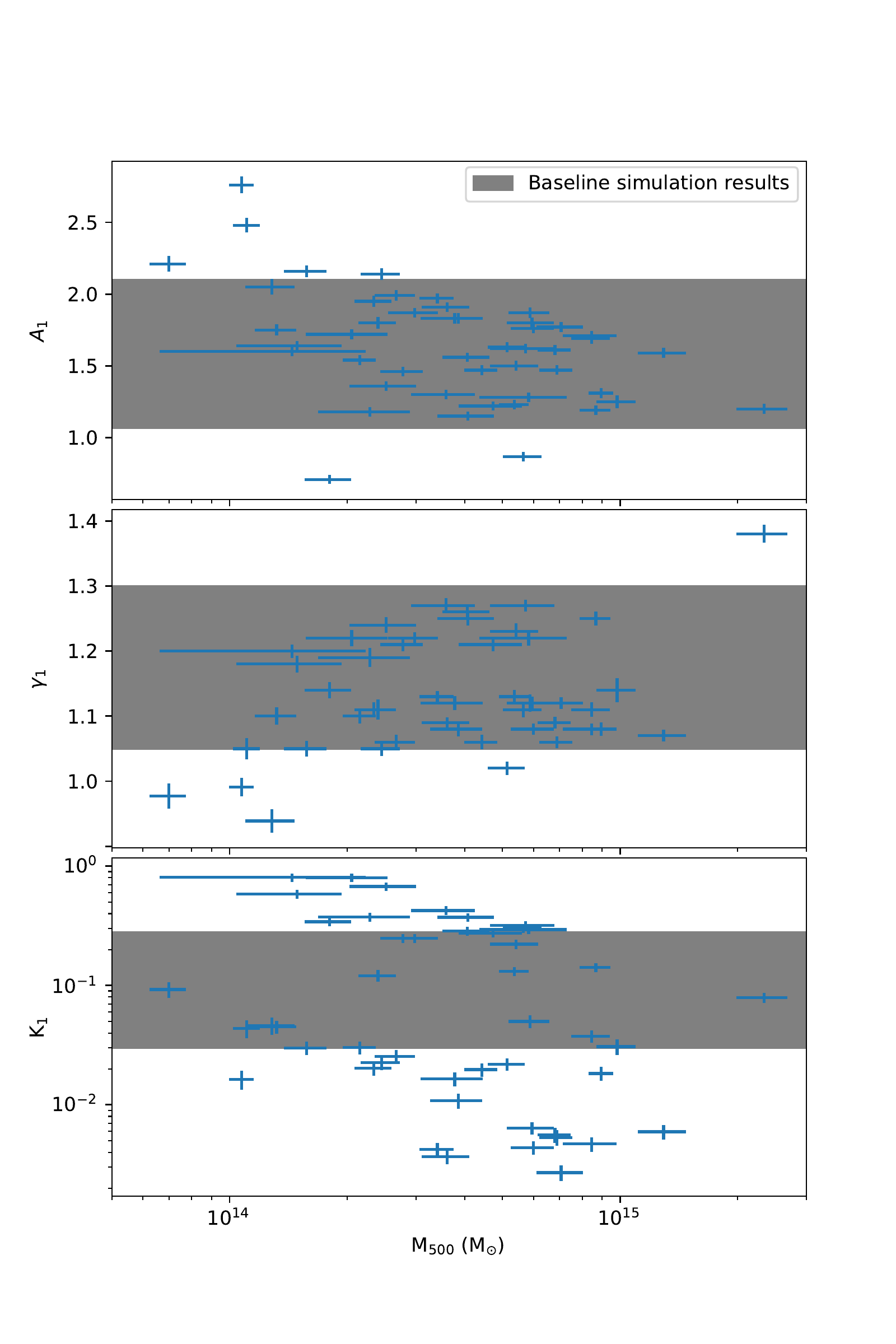}
	\caption{Best-fit parameters of the power-law entropy model $A_1$ (upper panel), $\gamma_1$ (middle panel), and $K_1$ (bottom panel) with their respective errors as functions of $M_{500}$. The shaded regions show the predictions from baseline simulations (e.g., \citealt{tozzi01}; \citealt{voit05b}).\label{fig:pl-p}}
\end{figure}

To quantify the statistical significance of the result that best-fit entropy profiles generally converge asymptotically to the baseline profile near the virial radius, we perform the $\chi^2$ fitting to the observed entropy profile ($K_{\rm obs}$) with the power-law model (\citealt{voit05b}), i.e.,
\begin{equation}\label{eq:powerlaw}
	\frac{K_{\rm obs}}{K_{200}}=K_1+A_1\times(\frac{r}{r_{200}})^{\gamma_1},
\end{equation}
where $\gamma_1$ is the slope of the entropy profile, $A_{1}$ is the normalization, and $K_1$ is the scaled central entropy. In order to compare the best-fit parameters directly to the baseline prediction, we define the scaling factor $K_{200}$ following \citet{voit05b}, i.e., 
\begin{equation}
K_{200}=362\ {\rm keV\ cm^2}\frac{T_{200}}{1\ {\rm keV}} \times \left[\frac{H(z)}{H_0}\right]^{-4/3}\left(\frac{\Omega_m}{0.3}\right)^{-4/3}, 
\end{equation}
where 
\begin{equation}
T_{200}\equiv \frac{GM_{200}\mu m_{\rm p}}{2r_{200}}.
\end{equation}
We find that for all sample members, the reduced $\chi^2$ is less than or approximately equal to 1, which implies that the power-law model is sufficient to describe the entropy profile. In Figure \ref{fig:pl-p} we plot the best-fit coefficients of the power-law model and their uncertainties for each sample member as a function of $M_{500}$. We find that $A_{1}$ and $\gamma_1$ generally fall in the predicted range of the baseline simulations (e.g., \citealt{tozzi01}; \citealt{voit05b}). $K_1$ on the other hand, diverts from the baseline prediction (see figure 5 of \citealt{voit05b}), implying that the feedback processes and the radiative cooling significantly affect the thermal properties of the ICMs in the core regions of galaxy clusters and groups.     
 
\section{Discussions} \label{sec:dis}
\subsection{Impact of Systematic Errors}\label{sec:err}
The results obtained with our model may be biased by systematic errors primarily including the uncertainties in calibrating instruments and the simplifications made in the calculations (e.g., \citealt{cava09}; \citealt{buote16}). In order to validate our results in the following subsection we will focus on eight error sources that exist in measuring gas temperature and X-ray surface brightness; actually the errors caused by these sources have been taken into account in the fittings in Section \ref{sec:mcmc}.
\subsubsection{Systematic Errors in Measuring Gas Temperature}\label{sec:err-temp} 
\noindent\textbf{Instrument calibration:}
Primarily due to the energy-dependent difference in effective areas between the X-ray instruments on board \Chandra{}, \XMM{} and \Suzaku{}, even after careful calibration the gas temperatures measured with these instruments always exhibit differences, the level of which may depend on the temperature of the target. As pointed out in \citet{kett13} and \citet{S15}, the difference of the temperatures measured with \Chandra{} and \XMM{} can be about $7\%$, $16\%$, and $23\%$ for targets with averaged temperature $2$, $5$, and $10$ keV, respectively, while the difference between the \XMM{} and \Suzaku{} measurements is about $\sim 5\%$, which is less temperature dependent. In this work we evaluate the impact of this effect by following the method presented in Z16, i.e., we add a systematic error $\Delta_{\rm inst,T}$ to the observed temperature (\citealt{S15}; \citealt{kett13}), which is given by $\Delta_{\rm inst,T}=(T-T^{0.889})/T$.

\noindent\textbf{Thermal plasma models and atomic database:}
The uses of different thermal plasma models and different atomic databases may cause systematic errors in the fittings. Previous studies (e.g., \citealt{matsushita03}; \citealt{sato11}) show that using two sets of the most popular atomic codes/tables, i.e., \texttt{AtomDB}\footnote{http://www.atomdb.org/} (embedded in the \texttt{APEC} model) and \texttt{SPEXACT}\footnote{https://www.sron.nl/astrophysics-spex} (embedded in the \texttt{CIE} model, which is updated from the \texttt{MEKAL} model) will result in a limited difference in gas temperature measurements, which is, however, typically smaller than the statistics error.
\citet{mernier19} have compared these two models by simulating fake spectra with one of them and then fit the spectra with the other, and found that the temperatures obtained with the two models show a difference of $\sim 0.05$ keV when $T < 1$ keV, or a difference of $\lesssim 5\%$ when $T \ge 1$ keV. Therefore, we have added a systematic error $\Delta_{\rm atom,T}=0.05$ KeV if $T<1$ keV or $\Delta_{\rm atom,T}=0.05T$ if $T\ge 1$ keV to the observed temperature.   

\noindent\textbf{Metal abundance in outer regions:}
Due to the limited S/N, the abundance of the outer regions cannot be tightly constrained in many cases. Thus it is often fixed to typical values such as $0.3$ solar or to the abundance of the adjacent inner region. Since the measurements of metal abundance and gas temperature are coupled, the uncertainties in the measurement of the abundance will be transferred to the measurement of gas temperature, which is typically $\lesssim 3\%$ as found by \citet{vikhlinin05} (see also \citealt{su15}; \citealt{lak16}). We have added the error $\Delta_{\rm Aout,T}=0.03T$ accordingly to the observed temperature of the corresponding outer regions.

\noindent\textbf{Non-equilibrium between electron and proton populations:}
As proposed by \citet{hoshino10}, it is possible that the electrons may not be in thermal equilibrium with the protons near the virial radius, which will introduce a systematic error if the thermal equilibrium state is assumed in the model. Currently, no firm observational evidence has been presented to support this idea, and it is estimated that within $r_{100}$ the difference between electron and proton temperature is less than $1\%$ \citep{wong09}. Therefore we add the systematic error $\Delta_{\rm pe,T}=0.01T$ to the temperatures measured outside $r_{500}$. 

\noindent\textbf{Calculation of 2-dimensional temperature:}
In Section \ref{sec:model} we have modeled the 2-dimensional temperature using the method of \citealt{mazza04} (Equation \ref{Eq:T2d}) in the RTI model fitting. This method, however, will lead to at most $10\%$ systematic errors \citep{mazza04}. We then add $\Delta_{\rm 2d,T}=0.1T$ to the 2-dimensional temperatures.
 
\noindent\textbf{Possible multi-phase gas in the central region:}
In the cases when a single-phase temperature model is used (Section \ref{sec:Chandra-data-analysis}), it is possible that there exists an unresolved cool phase gas, the absence of which may cause a small systematic error in the spectral fitting. In order to estimate the possible model bias in such cases, we use the \texttt{XSPEC} command \texttt{fakeit} to create a test spectrum that consists of two \texttt{APEC} components. The temperature of the cool component is set to be $0.8$ keV, while that of the hot phase is set to be $2.0$ keV, $5.0$ keV, and $10.0$ keV, respectively. The abundances of both phases are set to be $0.4$ solar, and the normalization of the cool component is determined in such a way that the cool phase accounts for $3\%$ of the total emission. 
We fit the test spectrum with a single-APEC model and find that in all cases the systematic errors less than $1\%$ in measuring the temperature of the single-APEC model  will arise when the cool phase component is ignored in the spectral fitting. Thus, we have decided to add an additional systematic error $\Delta_{\rm multi,T}= 0.01 T$ to the ICM temperature measured within innermost $50$ kpc when the single temperature model is applied.

\subsubsection{Systematic Errors in Measuring X-ray Surface Brightness}\label{sec:err-sbp}
\noindent\textbf{Instrument Calibration:}
As shown in \citealt{S15} and \citealt{kett13}, for a given target the difference of the X-ray surface brightness measured via \Chandra{}, \XMM{} and \Suzaku{} can be up to about $10\%$ due to the effective area calibration uncertainties. Therefore, we have added a systematic error $\Delta_{\rm inst,S}=0.1S_{\rm obs}$ to the observed surface brightness. 

\noindent\textbf{Calculation of cooling function:}
The gas emissivity depends linearly on the cooling function (Equation \ref{sbp}), which further depends on the gas metal abundance. Systematic errors will rise if the metal abundances cannot be well constrained in the outer regions (Section \ref{sec:err-temp}). By altering the abundance in the typical range ($0.1$ to $0.5$; \citealt{lovisari19}) we find that the cooling function is shifted by about $5\%$. Hence we adopt a systematic error $\Delta_{\rm abund,S}=0.05S_{\rm model}$ to the calculated surface brightness profile to account for this effect.

\subsection{The Necessity of RTI Model parameters}\label{sec:model-comp}
\begin{deluxetable}{cccccl}
	\centering
		\tablecaption{Sample averaged $R_{\rm eff}$ for cases tested in Section \ref{sec:model-comp}\tablenotemark{a} \label{tab:reff_comp}}
		\tablehead{\colhead{Group A}& \colhead{Group B} & \colhead{Group C} & \colhead{Group D} & \colhead{Group E} & \colhead{$R_{\rm eff}$} \\
			\colhead{$c_1$ to $c_5$}& \colhead{$A$, $B$, $\gamma$} & \colhead{$r_{\rm shock}$, $\Gamma$} & \colhead{$N_2$, $N_3$, $C_{\rm w}$} & \colhead{$\rho_0$, $r_s$, $\delta_1$, $\delta_2$, $A_0$, $\gamma_0$, $K_0$, $f_{\rm g}$} & \colhead{}}
		\startdata
		& 		& 		& 		& $\checkmark$	& -1.20 (Case 5)\\
		$\checkmark$ 	& 		& 		& 		& $\checkmark$ 	&-0.22 \\
		& $\checkmark$	& 		&	 	& $\checkmark$ 	&-0.30 \\
		&	  	& $\checkmark$  & 		& $\checkmark$ 	&-0.31 \\
		& 		& 		& $\checkmark$ 	& $\checkmark$	& 0.20 (Case 4) \\
		$\checkmark$	& $\checkmark$	& 		& 		& $\checkmark$	&-0.18 \\
		$\checkmark$ 	& 		&$\checkmark$ 	& 		& $\checkmark$	& -0.18\\
		$\checkmark$	& 		& 		& $\checkmark$	& $\checkmark$	& 0.21 \\
		& $\checkmark$	& $\checkmark$	& 		& $\checkmark$	& -0.12\\
		& $\checkmark$ 	& 		& $\checkmark$	& $\checkmark$ 	& 0.17\\
		& 		& $\checkmark$	& $\checkmark$ 	& $\checkmark$ 	& 0.45 (Case 3) \\
		$\checkmark$ 	& $\checkmark$ 	& $\checkmark$	& 	 	& $\checkmark$ 	& 0.31\\
		$\checkmark$ 	& $\checkmark$ 	& 	 	& $\checkmark$	& $\checkmark$ 	& 0.28\\
		& $\checkmark$ 	& $\checkmark$ 	& $\checkmark$ 	& $\checkmark$ 	& 0.51 (Case 2) \\
		$\checkmark$ 	& $\checkmark$ 	& $\checkmark$ 	& $\checkmark$ 	& $\checkmark$ 	& 0.53 (Case 1)\\
		\enddata
		\tablenotetext{a}{Parameters that were included and set free in the fitting are marked by $\checkmark$ for each case.}
\end{deluxetable}
An empirical description of gas temperature and surface brightness requires about ten free parameters ($\sim 4-7$ for temperature, e.g., \citealt{allen01}; \citealt{zhang06}; \citealt{vikhlinin06}; $\sim 3-6$ for the $\beta$ or double-$\beta$ model of the surface brightness), although the RTI model as well as some previous studies (e.g., \citealt{ostriker05}; \citealt{patej15}; Z16) demonstrated the possible necessity to include more free parameters for describing the physical processes behind. 
In order to evaluate whether all the parameters listed in Table \ref{tab:prior} are necessary for the current RTI model, we have compared the best-fit model predictions obtained from the RTI model fitting in the following five representative cases: 
(1) best-fit results as given in Section \ref{sec:mcmc} [21 free parameters]; (2) five parameters related to the gas clumping profile (i.e., $c_1$, $c_2$, $c_3$, $c_4$, and $c_5$; parameter group A) are neglected in the fitting, and $C(r)$ are set to be $1$ [16 free parameters]; (3) in addition to case 2, three parameters related to the non-thermal pressure (i.e., $A$, $B$, and $\gamma$; parameter group B) are neglected in the fitting, and $\eta(r)$ are set to be $1$ [13 free  parameters]; (4) in addition to case 3, values of two parameters related to the accretion-shock (i.e., $r_{\rm shock}$ and $\Gamma$; parameter group C) are fixed to the averaged values derived from the observation of \citet{patej15} [11 free parameters]; (5) in addition to case 4, values of three parameters related to the energy conservation (i.e., $N_2$, $N_3$, and $C_{w}$; parameter group D) are fixed to the theoretical or simulated values that have been accepted and used among astronomical community (e.g., \citealt{cnm12}; \citealt{nel14}; Z16) [8 free parameters; parameter group E].

\begin{startlongtable}
	\begin{deluxetable}{lccccc}
		\tablecaption{AIC values for the model predictions obtained in the five cases described in Section 5.2. \label{tab:aic}}
		\tablehead{\colhead{Name\tablenotemark{a}} & \colhead{Case 1} & \colhead{Case 2} &\colhead{Case 3} &\colhead{Case 4} &\colhead{Case 5} }
		\startdata
1E 1455.0+2232 & 2351.8 & 2326.8 & 2426.8 & 2490.7 & 2427.5 \\
Abell 1068 & 2403.3 & 2384.5 & 2414.6 & 2620.9 & 3283.4 \\
Abell 1246\tablenotemark{1} & 2515.6 & 2600.8 & 2659.7 & 2684.7 & 2747.4 \\
Abell 13 & 2336.0 & 2340.2 & 2348.7 & 2335.6 & 2431.5 \\
Abell 133\tablenotemark{1} & 1876.0 & 1886.2 & 1899.4 & 1900.5 & 2560.4 \\
Abell 1413 & 2149.7 & 2144.9 & 2160.6 & 2240.2 & 2261.4 \\
Abell 1689 & 2583.3 & 2540.9 & 2584.7 & 2577.4 & 2832.3 \\
Abell 1775 & 2276.2 & 2242.5 & 2268.1 & 2282.4 & 2386.2 \\
Abell 1795\tablenotemark{1} & 1616.0 & 1631.2 & 1635.2 & 1761.2 & 2140.5 \\
Abell 1835\tablenotemark{1} & 2518.2 & 2526.2 & 2526.7 & 2710.0 & 2952.5 \\
Abell 2029\tablenotemark{1} & 2388.5 & 2398.3 & 2423.3 & 2514.3 & 2783.6 \\
Abell 209\tablenotemark{1} & 2361.5 & 2390.4 & 2397.0 & 2425.0 & 2437.3 \\
Abell 2142\tablenotemark{1} & 1097.2 & 1169.2 & 1202.5 & 1220.9 & 1569.0 \\
Abell 2163 & 2464.5 & 2440.0 & 2450.3 & 2463.1 & 2469.5 \\
Abell 2199 & 2230.6 & 2223.7 & 2269.9 & 2307.6 & 3124.1 \\
Abell 2204\tablenotemark{1}  & 2243.7 & 2250.1 & 2272.2 & 2412.8 & 2638.5 \\
Abell 2255\tablenotemark{1} & 2493.1 & 2562.1 & 2579.8 & 2606.0 & 2683.6 \\
Abell 2319\tablenotemark{1} & 1584.8 & 1631.0 & 1630.1 & 1626.0 & 1712.4 \\
Abell 2597 & 2325.5 & 2323.4 & 2383.0 & 2598.9 & 3183.7 \\
Abell 2667 & 2366.9 & 2350.3 & 2384.7 & 2543.1 & 2548.4 \\
Abell 3158 & 2541.7 & 2539.4 & 2576.7 & 2594.1 & 2623.7 \\
Abell 3266 & 2548.4 & 2535.6 & 2552.8 & 2568.7 & 2572.9 \\
Abell 383 & 2294.7 & 2291.4 & 2337.4 & 2325.9 & 3066.6 \\
Abell 478 & 2102.1 & 2064.1 & 2102.7 & 2209.1 & 2266.2 \\
Abell 644\tablenotemark{1} & 2471.6 & 2557.5 & 2621.3 & 2701.7 & 2854.1 \\
Abell 68\tablenotemark{1} & 2418.9 & 2461.5 & 2482.3 & 2516.2 & 2510.5 \\
Abell 773 & 2465.0 & 2451.8 & 2486.7 & 2507.1 & 2526.0 \\
Abell s1101\tablenotemark{1} & 2213.7 & 2244.4 & 2324.1 & 2613.4 & 2904.6 \\
Centaurus cluster & 2164.6 & 2170.1 & 2176.3 & 2184.4 & 3283.0 \\
Cl 0016+16\tablenotemark{1} & 2127.7 & 2150.1 & 2122.4 & 2130.6 & 2125.5 \\
Cl 0024+17 & 2287.5 & 2288.0 & 2307.2 & 2307.6 & 2306.8 \\
Coma cluster & 1592.6 & 1548.7 & 1609.4 & 1656.0 & 1669.2 \\
ESO 306- G 017 GROUP & 2402.5 & 2374.9 & 2412.0 & 2472.6 & 2532.2 \\
HydraA Cluster & 2182.3 & 2148.5 & 2176.7 & 2241.4 & 2195.1 \\
Perseus cluster\tablenotemark{1} & 2104.1 & 2259.0 & 2279.3 & 2355.8 & 2430.3 \\
Pks 0745-191 cluster\tablenotemark{1}  & 2338.8 & 2345.4 & 2392.7 & 2662.8 & 3247.9 \\
RXC J0605.8-3518 & 2371.2 & 2363.4 & 2400.1 & 2508.0 & 2760.9 \\
Rxc j1825.3+3026 & 2578.5 & 2578.9 & 2600.2 & 2600.7 & 2625.1 \\
RXC J2234.5-3744 & 2429.4 & 2403.6 & 2447.1 & 2392.4 & 2468.8 \\
Rx j1120.1+4318\tablenotemark{1} & 1958.2 & 2036.1 & 2000.9 & 2067.8 & 2011.2 \\
Rx j1159.8+5531 & 2439.4 & 2411.6 & 2417.8 & 2393.1 & 2720.0 \\
Rx j1334.3+5030 & 2234.7 & 2232.4 & 2235.8 & 2285.7 & 2316.0 \\
RX j1347.5-1145\tablenotemark{1} & 2512.1 & 2555.6 & 2591.1 & 2910.0 & 3188.8 \\
Ugc 03957 cluster & 2348.9 & 2336.3 & 2364.0 & 2391.1 & 2524.1 \\
Virgo cluster\tablenotemark{1} & 2305.2 & 2316.7 & 2367.6 & 2328.9 & 2446.9 \\
Zwcl 1215.1+0400 & 2527.4 & 2533.4 & 2575.9 & 2591.2 & 2621.5 \\
ZWCL 3146 & 2442.6 & 2439.4 & 2466.2 & 2848.0 & 3259.3 \\
		\enddata
		\tablenotetext{a}{The superscripts 1 indicate that the effect of gas clumping is important and can significantly improve the fittings by including it in the model.}
	\end{deluxetable}
\end{startlongtable}

According to the definition of model efficiency, larger value of $R_{\rm eff}$ indicates better model fit within the range of $(-\infty, 1]$. An acceptable model fit is acquired when $R_{\rm eff} \gtrsim 0$, and the model best describes the observation when $R_{\rm eff}=1$ is achieved (e.g., \citealt{nash70}; \citealt{engeland02}).
The derived sample-averaged $R_{\rm eff}$ (Tale \ref{tab:reff_comp}) for the five cases are 0.53, 0.51, 0.45, 0.20, and -1.2, respectively, where cases 1 to 4 gives acceptable fits and case 1 best describes the observation among the five cases. In addition, we have also tested cases given by altering the order of neglecting or fixing to the parameters described in cases 2 to 5, and all of them yielded averaged $R_{\rm eff}$ smaller than 0.53 (Table \ref{tab:reff_comp}), the averaged $R_{\rm eff}$ of the original best-fit in Section 4, suggesting that among all cases, the original model (case 1) as introduced in Section 4 best describes the observation. The most consequential parameter group that improves $R_{\rm eff}$ is the energy conservation parameters (group D), and the minimum number of parameters needed to achieve $R_{\rm eff} \gtrsim 0$ is 11, as used in Case 4.
	
To give a further evaluation on the statistical significance, we 
used the Akaike Information Criterion (AIC; \citealt{akai74}) to decide which one best describes the observations. The AIC, which has been widely used as a model selection criterion (e.g., \citealt{aho14}), is defined as 
\begin{equation}
AIC=2k-2\ln(L),
\end{equation}
where $k$ is the total number of all model parameters, and $L$ is the likelihood function that is defined to characterize the goodness of fit of the model to the observed data (Equation \ref{eq:lhood}). Given two sets of model predictions, the relative likelihood $L_{R}$ of them is calculated as $\exp((AIC_{1}-AIC_{2})/2)$ and is used in the likelihood-ratio test to judge which prediction is better. In the likelihood ration test we set the threshold p-value to be $0.05$, a value commonly used in astronomical literature for the significance test. When $L_{R} < 0.05$, which corresponds to  $AIC_{1}-AIC_{2}\lesssim -6$, the model prediction with a smaller AIC value is regarded as the better one to describe the dataset. Otherwise neither prediction can be regarded better than the other. In Table \ref{tab:aic} we list the AIC values calculated in the above five cases for all sample members. 
The sample averaged AIC value of the five cases are 2268, 2277, 2305, 2376, and 2579, respectively. The AIC differences between cases 2 to 5 and case 1 are -9, -37, -108, and -311 respectively, providing statistical evidence for case 1's superiority over cases 2 to 5. Therefore, albeit the seeming large number of free parameters (e.g., \citealt{burnham02}; \citealt{claeskens16}), the best-fit model in Section 4 can be considered as the appropriate one to describe the observation, suggesting that all the RTI model parameters are necessary in order to describe the corresponding physical processes in the fitting (Section \ref{sec:mcmc}).

\subsection{The Impact of Gas Clumping}\label{sec:gas-clumping}
\subsubsection{Gas Clumping Signal in the Center of Galaxy Clusters and Groups}
	\begin{figure}
		\plotone{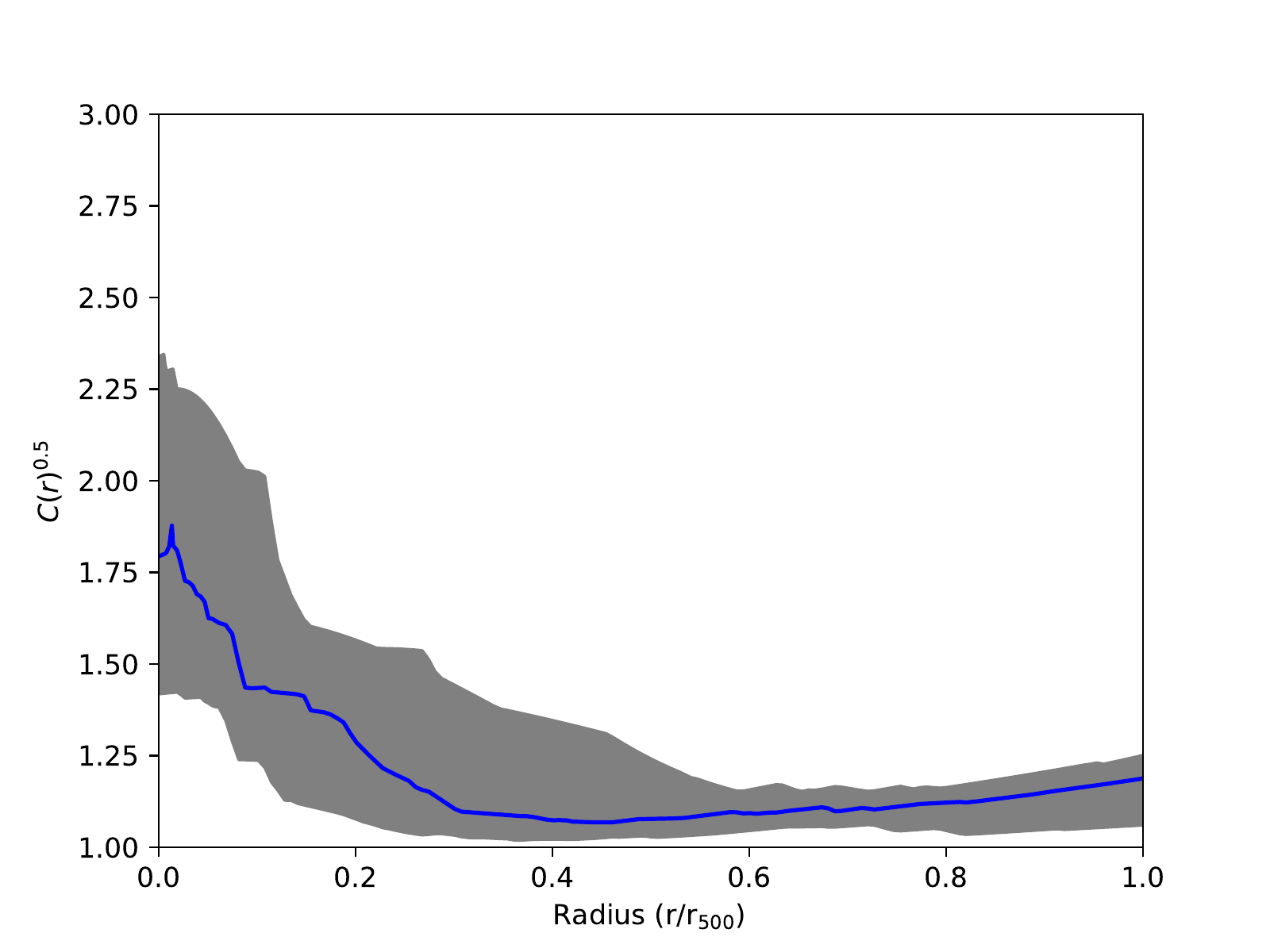}
		\caption{Averaged clumping factor as a function of radius for all sample members. The shaded area represents the $68\%$ uncertainty range.\label{fig:clumping_factor}}
	\end{figure}
\begin{figure}
	\centering
	\includegraphics{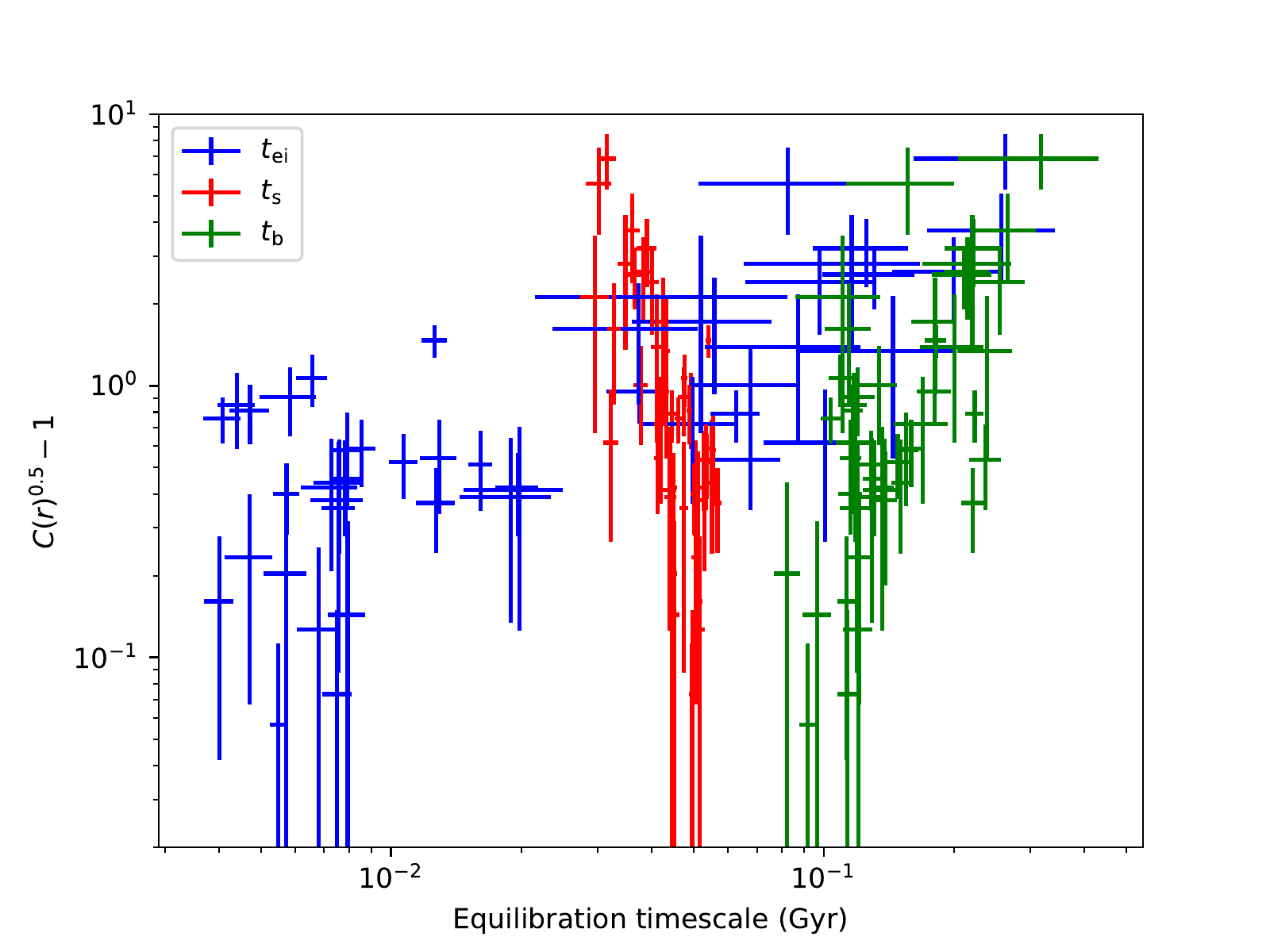}
	\caption{Gas clumping factor as a function of the electron-ion equilibration (blue; $t_{\rm ei}$), sound-crossing (red; $t_{\rm s}$), and buoyancy (green; $t_{\rm b}$) time scale at $0.05$ $r_{500}$ for all sample members.\label{fig:tei}}
\end{figure}

The sample-averaged best-fit clumping profile has a significant signal inside $\sim 0.1 r_{500}$ (see Figure \ref{fig:clumping_factor}) as expected from cosmological simulations (e.g., \citealt{vazza13}) and found in the observations of Perseus and Coma clusters (e.g., \citealt{chura12}; \citealt{zhurav15}). Since such clumping signals appear in the simulation when effects of the AGN activity and the radiative cooling are taken into account, we propose that accumulated effects of the AGN historical feedback and the radiative cooling have contributed to the central clumping signal. In order to verify the assumption, we calculate the electron-ion equilibrium time scale ($t_{\rm ei}$), the sound-crossing time scale ($t_{\rm s}$), and the buoyancy time scale ($t_{\rm b}$) at $0.05$ $r_{500}$ of sample members, the longest of which can be used as an indicator of the relaxation time for the gas at the cluster center.  Following \citet{hoshino10}, we calculate $t_{\rm ei}$ as 
	\begin{equation}
	t_{\rm ei}(r) \simeq 2.0\times 10^8\ {\rm yr}\frac{(T(r)/10^8\ {\rm K})^{3/2}}{(n_{\rm p}(r)/10^{-3}\ \rm{cm^{-3}})(ln \Lambda_{c}(r)/40)},
	\end{equation}   
	where $\ln\Lambda_{c}(r)$ is the Coulomb logarithm, which is calculated as
	\begin{equation}
	\ln\Lambda_{c}(r)=30-\ln(\frac{n_{\rm e}(r)}{{\rm cm^{-3}}}^{1/2}\frac{T(r)}{\rm eV}^{-3/2}).
	\end{equation}
	$t_{\rm s}$ and $t_{\rm b}$ are calculated following \citet{gitti12} as $t_{\rm s}(r)=r/\sqrt{5k_{\rm b}T(r)/3\mu m_{\rm p}}$ and $t_{\rm b}(r)=r/\sqrt{2GM(<r)R_{\rm cs}/r^2}$, where $R_{\rm cs}$ is the scale of gas clumping at $0.05$ $r_{500}$ and is estimated to be 10 kpc.	  
	We plot $t_{\rm ei}$, $t_{\rm s}$, and $t_{\rm b}$ as functions of the clumping factor ($C(r)^{0.5}-1$) at $0.05$ $r_{500}$  for each sample member in Figure \ref{fig:tei}. The correlations between the three time scales ($t_{\rm ei}$, $t_{\rm s}$, and $t_{\rm b}$) and the clumping factor are $0.82 \pm 0.07$, $-0.62 \pm 0.06$, and $0.70 \pm 0.09$, respectively. As shown in Figure \ref{fig:tei}, $t_{\rm b}$ is the largest out of the three time scales, which is on the order of AGN cycling time ($\sim 10^7-10^8$ yr; \citealt{blanton10}), and sample members with longer $t_{\rm b}$ generally have larger clumping factors at $0.05$ $r_{500}$. This result supports our argument that the accumulated effect of AGN historical feedback have contributed to the central clumping signal since it will be more difficult for the sample member with a larger equilibrium time to become relaxed after the AGN activity disturbs the core region. 

\subsubsection{Is Gas Clumping Effect Common in Clusters and Groups?}
Cosmological simulations of, e.g., \citet{nagai11} and \citet{vazza13}, predicted that at cluster or group outskirts the ICM is inhomogeneous and apparently deviates from the hydrostatic equilibrium. In the past decade corresponding studies, which mainly focused on the effects gas clumping, have been conducted based on the observations of \Chandra{} (e.g., \citealt{chura12}; \citealt{zhurav15}), \ROSAT{} (e.g., \citealt{eckert12}; \citealt{eckert15}), and \XMM{} (e.g., \citealt{ghirar18}; \citealt{eckert19}). 
In order to investigate what is the major difference between the targets showing clear evidence for gas clumping and those not, and, from another perspective, to answer the question whether or not the gas clumping effect is actually common in clusters and groups, we have divided the targets into two corresponding subsamples, one (subsample A) containing 19 targets (8 of which are included in W12a sample) show significant improvement when the effect of gas clumping is considered in the model (marked with the superscript "1" in Table \ref{tab:aic}), and the other (subsample B) containing the rest 28 targets.

\begin{figure}
	\plotone{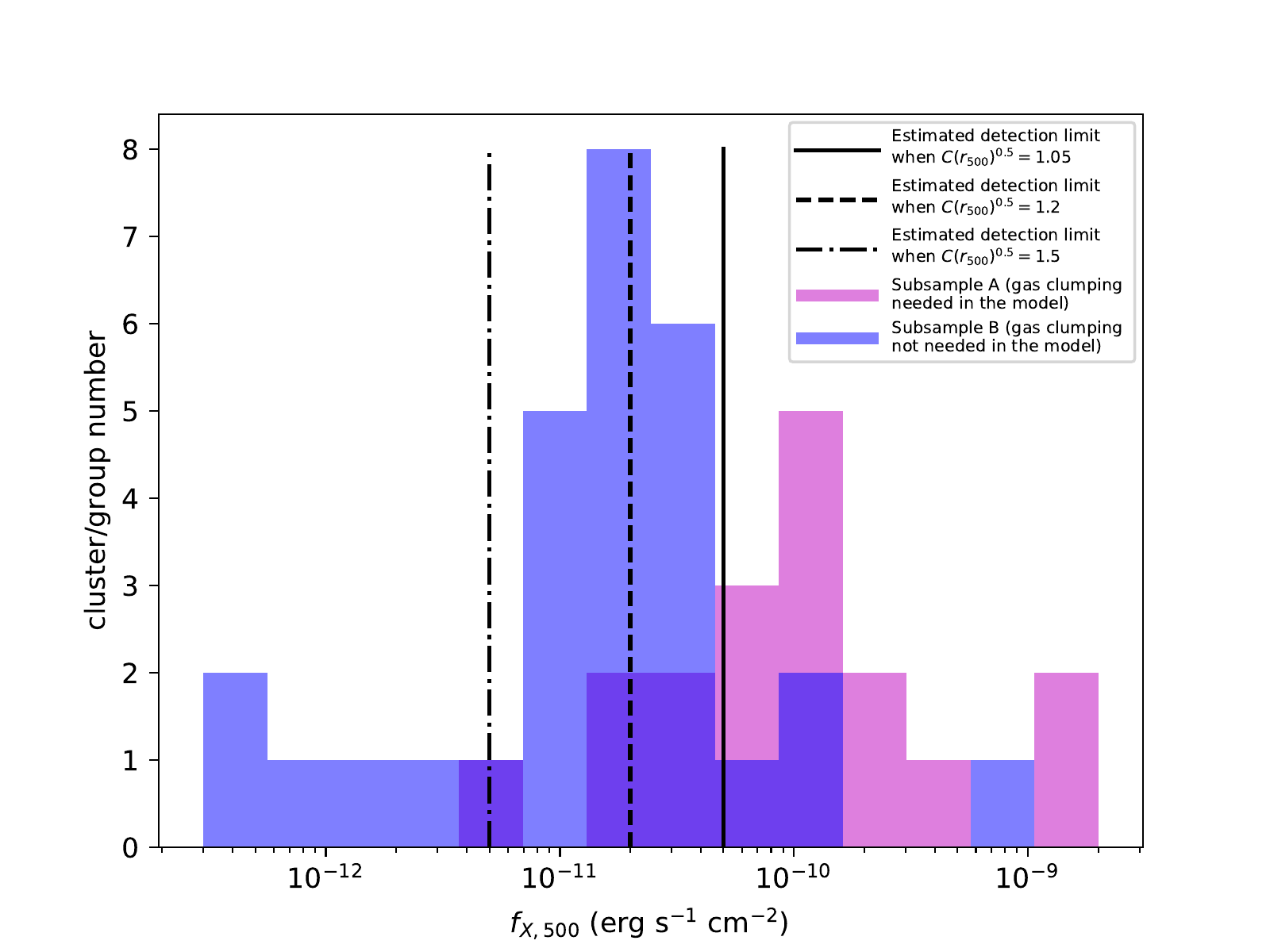}
	\caption{Number distributions of clusters and groups as a function of $f_{X,500}$. Vertical lines show the estimated detection limit for different clumping levels, which are characterized by the clumping factor $C(r_{500})$ (see Figure \ref{fig:clumping_factor}). A $50$ ks observations is assumed for current X-ray missions such as \Chandra{} or \XMM{}.\label{fig:fx-dist}}
\end{figure} 
We have attempted to compare the redshifts, the gas temperatures averaged between $0.2-0.5$ $r_{500}$, the virial masses, the surface brightness concentrations (i.e., the integrated surface brightness within $0.2r_{500}$ divided by that within $r_{500}$), and the $0.7-7.0$ keV fluxes integrated inside $r_{500}$ ($f_{X,500}$) of the two subsamples by applying the Kolmogorov–Smirnov test (e.g., \citealt{naess12}). In particular, for each of the physical properties we calculate the empirical cumulative distribution functions  (ECDFs; e.g., \citealt{vaart98}) $F(x)$ calculated for two subsamples and use them to determine the maximum difference ($D_{n_{1},n_{2}}$), which is defined as
\begin{equation}
D_{n_{1},n_{2}}=\max_{x}{\abs{F_{n_{1}}(x)-F_{n_{2}}(x)}},
\end{equation}
where $n_1$ and $n_2$ are the sizes of the two subsamples, respectively, and $\max\limits_{x}$ represents the maximum value of the function $\abs{F_{n_{1}}(x)-F_{n_{2}}(x)}$ (e.g., for gas temperature) over the domain $x$. Thus we may conclude that, at the level of $\alpha$, where $\alpha=2\exp(-2c(\alpha)^2)$ and $c(\alpha )=D_{n_1,n_2}/\sqrt{\frac{n_1+n_2}{n_1n_2}}$, the hypothesis that the two subsamples possess the same properties is rejected (e.g., \citealt{knuth97}). Following \citealt{szy15}, when $\alpha$ is found to be less than the threshold value $0.05$ the two subsamples are regarded to be significantly different. We find that only $f_{X,500}$ shows a significant difference between the two subsamples ($\alpha=8.0\times 10^{-5}$) under this criteria. As shown in Figure \ref{fig:fx-dist}, where the distributions of target number as a function of $f_{\rm X,500}$ is plotted, the median values are $f^{A}_{{\rm X},500}=1.5 \times 10^{-10}$ $\rm erg\ s^{-1}\ cm^{-2}$ and $f^{B}_{{\rm X},500}=1.8 \times 10^{-11}$ $\rm erg\ s^{-1} cm^{-2}$ for the two subsamples, respectively.

For the current X-ray missions such as \Suzaku{}, \XMM{}, and \Chandra{}, the typical background level in $0.7-7.0$ keV is of the order of $10^{-6}$ $\rm photons\ cm^{-2}\ arcmin^{-2}\ s^{-2}$ (e.g., \citealt{hoshino10};  \citealt{nakajima18}). Meanwhile, the typical uncertainties in background modeling and instrument calibration reach about $10\%-20\%$ (e.g. \citealt{kushino02}; \citealt{gu16}) and $\sim 10\%$ (Section 5.1.2), respectively.  
These yield a detection threshold of $5\times 10^{-12}$, $2\times 10^{-11}$, or $5\times 10^{-11}$ $\rm erg s^{-1} cm^{-2}$ for target emission measured within $r_{500}$ to resolve the surface brightness increment caused by the gas clumping effect, if a typical 50 ks observation is performed and the clumping factor $C(r_{500})$ is set to 1.05, 1.2, or 1.5, respectively.
It is apparent that unless a target possesses a high $f_{{\rm X},500}$, which is typical for the targets in subsample B, it is impossible to reveal any information about gas clumping at $\sim r_{500}$, even if the clumps do exist. In clusters or groups with high X-ray fluxes, it seems that the gas clumping effect is very popular at $\sim r_{500}$.

\subsubsection{Detection of Gas Clumps in X-ray Maps}
It will be very interesting to investigate whether or not we are able to resolve gas clumps directly at $\sim r_{500}$ in the observed X-ray maps.
By studying the simulated X-ray maps \citet{vazza13} found that gas clumps exist on $\lesssim 50$ kpc scales in the outer regions ($\gtrsim r_{500}$) of all simulated galaxy clusters. Based on 21 \Chandra{} observations, \citet{zhurav15} confirmed the existence of such gas clumps in the central $220$ kpc of the Perseus cluster by analyzing the power spectrum of the X-ray maps. Can this phenomenon be directly observed at the cluster outskirts via imaging analysis? We have searched \Chandra{} archive\footnote{https://cxcfps.cfa.harvard.edu/cda/footprint/cdaview.html} for the clusters and groups satisfying the following three criteria: (1) the cluster or group should have been observed out to $r_{500}$ with a full or a nearly full azimuthal coverage; (2) the redshift should be lower than $0.3$ to enable the detection of gas clumps on $\lesssim 50$ kpc scales; (3) net photon counts inside a $50 \times 50$ $\rm kpc^{2}$ region should be more than 200 near $r_{500}$ to guarantee a sufficient sensitivity. We found only five clusters (i.e., Abell 133, Abell 1795, Abell 1882, Abell 1914, and Abell 2146) meet the first two criteria.  
Exposures of at least $1\times 10^6$ seconds at cluster outskirts are necessary, which are not currently available for all these clusters, to satisfy the third criterion. Clearly deep field observations are necessary in the future in order to resolve such gas clumps at $\sim r_{500}$.

\subsection{Comparison with Previous Works}\label{sec:comp-walker}
Our conclusion that the gas entropy profiles of the clusters in our sample are consistent with the power-law prediction of \citet{voit05b} also agrees with that of \citet{ghirar18}, who performed a joint X-ray and Sunyaev-Zel'dovich analysis for a sample of 12 galaxy clusters (10 of these clusters have been included in this work; see Table \ref{sample}), as well as those of, e.g., \citet{su15} and \citet{tcher16}, who carried out X-ray image spectroscopic studies on a single target. 
Apparently these results conflict with the conclusions drawn in a few other studies, such as W12a (all of 11 clusters are studied in this work). 
\begin{figure}
	\plotone{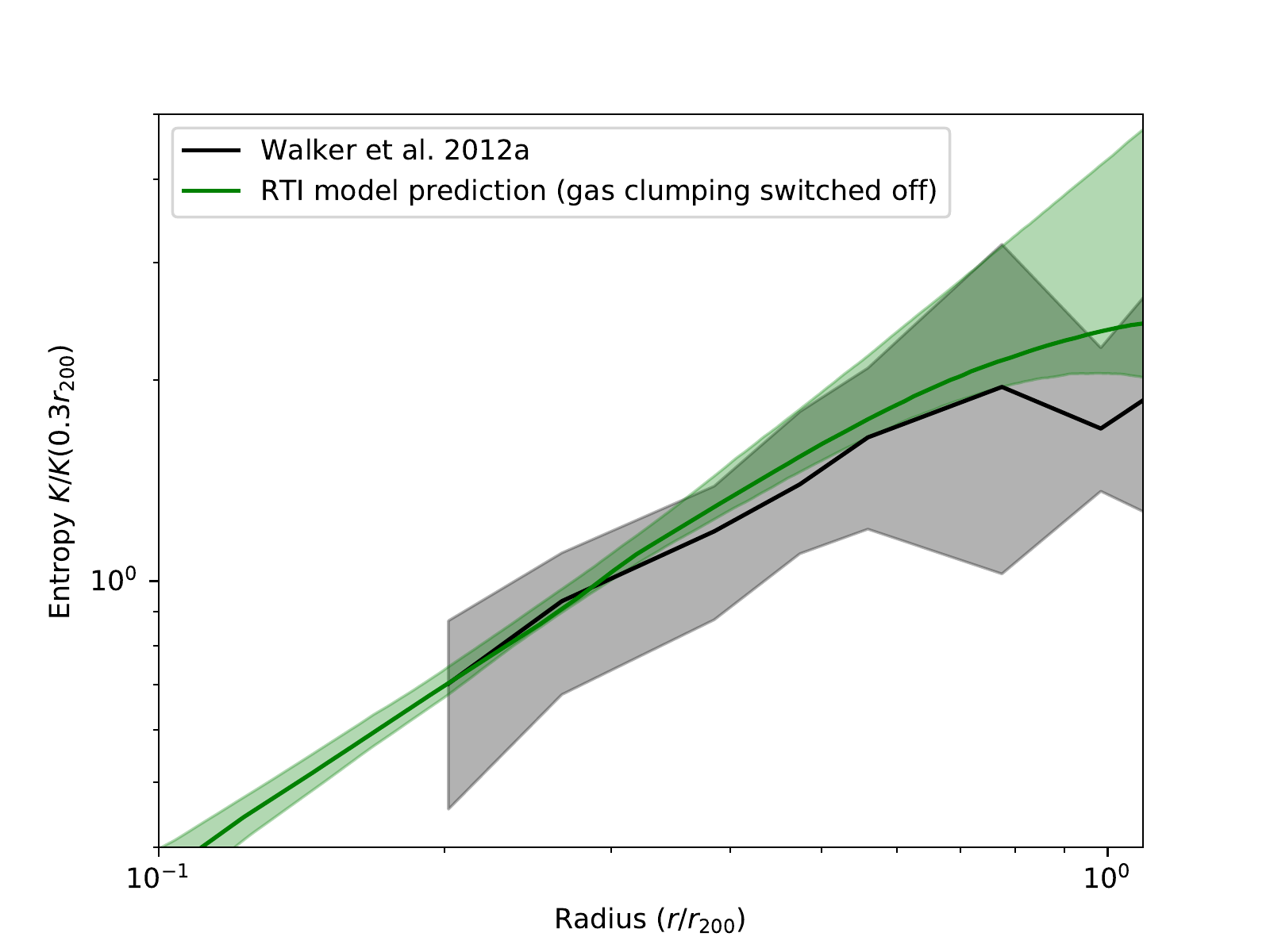}
	\caption{RTI model predicted gas entropy profile, which is averaged over 11 W12a clusters and calculated when the gas clumping effect is switched off (green line), along with that of W12a (black line). Shaded areas represent the $68\%$ uncertainty ranges.\label{fig:comp_walker}}
\end{figure} 
Given the fact that in our work the fittings of the gas temperature and X-ray surface brightness profiles among 8 out of 11 W12a clusters show significant improvement when the gas clumping effect is taken into account (see Table \ref{tab:aic}), it is reasonable to speculate that in W12a the gas density may have been overestimated at $\sim r_{500}$.
To verify this speculation we have tentatively rerun the RTI model fitting for the 11 W12a clusters by switching off the gas clumping effect (i.e., case 2 in Section \ref{sec:model-comp}). We find that the obtained gas distribution profiles and entropy profiles are consistent with those of W12a (68\% confidence level; Figure \ref{fig:comp_walker}).
In fact, W12a, \citet{ghirar18}, and other authors have suggested that it is very likely that the flattening of the entropy profiles will arise when the gas clumping effect is not properly considered in the model at $\sim r_{500}$. W12a also pointed out that neglecting the gas clumping effect in the model will cause the excess of gas fraction over the mean cosmic baryon fraction beyond $r_{500}$ (e.g., \citealt{simon11}).    

\subsection{Feedback Energy}
Using the derived gas entropy profiles we are able to study the energy injected into the ICM through the feedback processes. For a small gas element, the total feedback energy is $\Delta E_{\rm feed}(r)=\Delta E_{\rm heating}(r)+\Delta E_{\rm rad}(r)$, where $\Delta E_{\rm heating}(r)$ has been calculated with Equation \ref{E_H}, and $\Delta E_{\rm rad}(r)$ is the radiative loss that can be calculated as a time integration of the X-ray luminosity from $t_0$ (the age of the universe at $z=3$) to $t_{z0}$ (the age of the universe at the observation). In order to estimate $\Delta E_{\rm rad}(r)$ we employ the redshift-dependent mass-luminosity relations given in the simulation work of \citet{truong18} and the cluster mass evolution provided by \citealt{voit03} (see their figure 1), i.e., 
\begin{equation}
\Delta E_{\rm rad}(r)=\int^{t_{z_0}}_{t_0} f_L(m(t),t)\frac{L_X(r)}{f_L(m(t_{z_0}),t_{z_0})}dt,
\end{equation}  
where $f_L(m,t)$ denotes the luminosity-mass relation at time $t$, $m(t)$ is the corresponding cluster mass at time $t$ and is constrained by $m(t_{z0})=M_{500}$, and ${L_X}$(r) is the luminosity of the gas element at $z_0$
\begin{equation}
L_X(r)=C(r)n_e(r)n_p(r)\Lambda_{\rm bol}(T,Z)V^*,
\end{equation}
where $\Lambda_{\rm bol}(T,Z)$ is the cooling function in $0.1-50$ keV and $V^*$ is the volume of the gas element. 
\begin{figure}
	\plotone{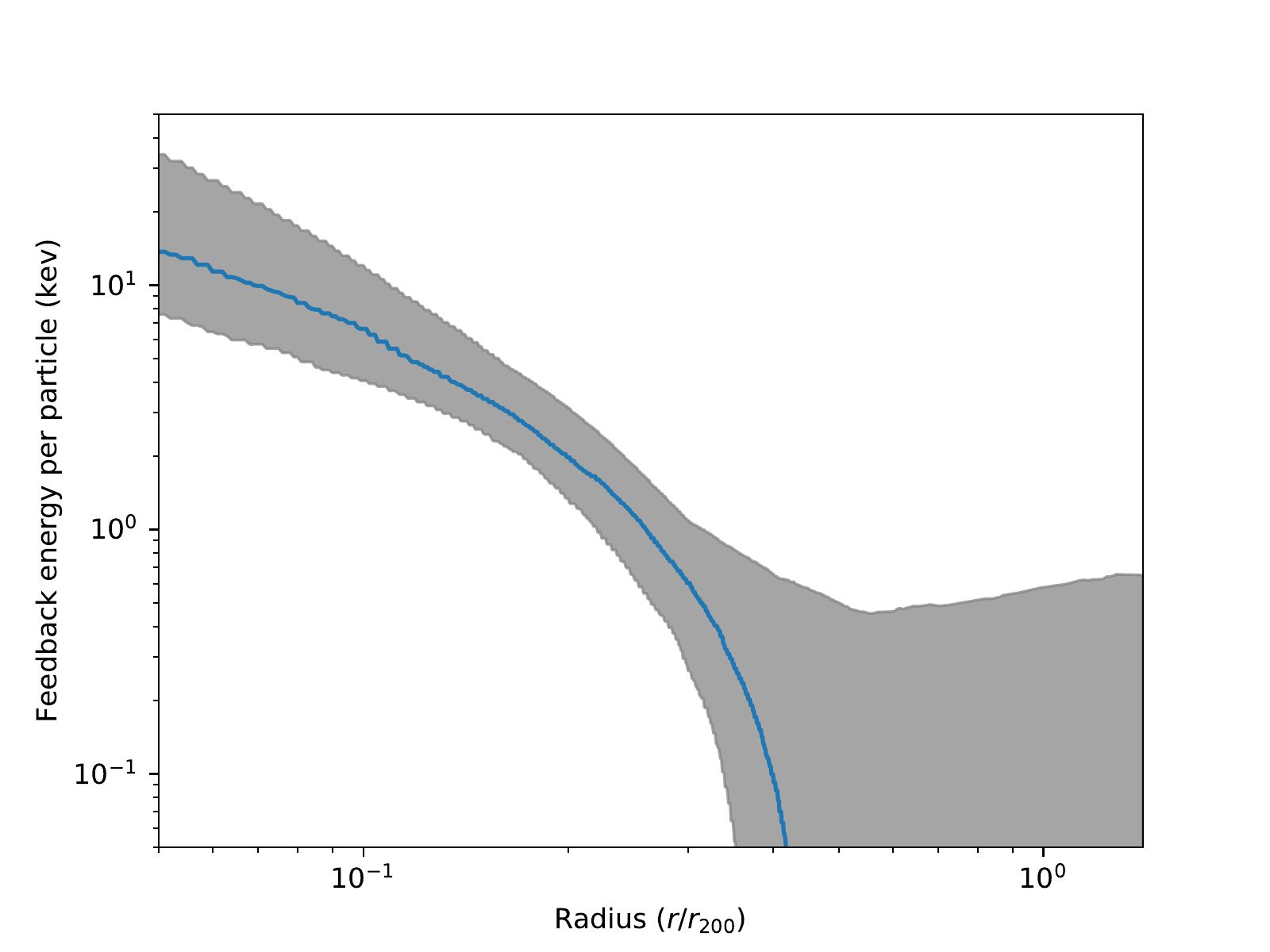}
	\caption{Sample averaged total feedback energy (blue line) with the $68\%$ uncertainty range (shaded region).\label{fig:feed-r}}
\end{figure}
We plot the calculated sample averaged feedback energy per gas particle as a function of radius in Figure \ref{fig:feed-r}, and find that the feedback energy outside $\sim 0.35$ $r_{200}$ is consistent with 0, suggesting that the pre-heating is likely to inject no more than 0.5 keV per particle (averaged between $0.35$ $r_{200}$ and $r_{500}$). Interestingly, the radius $0.35$ $r_{200}$ is similar to the radius within which a central abundance excess, if there exists one, is usually found (e.g., \citealt{lovisari19}, \citealt{makishima01}). 
\begin{figure}
	\plotone{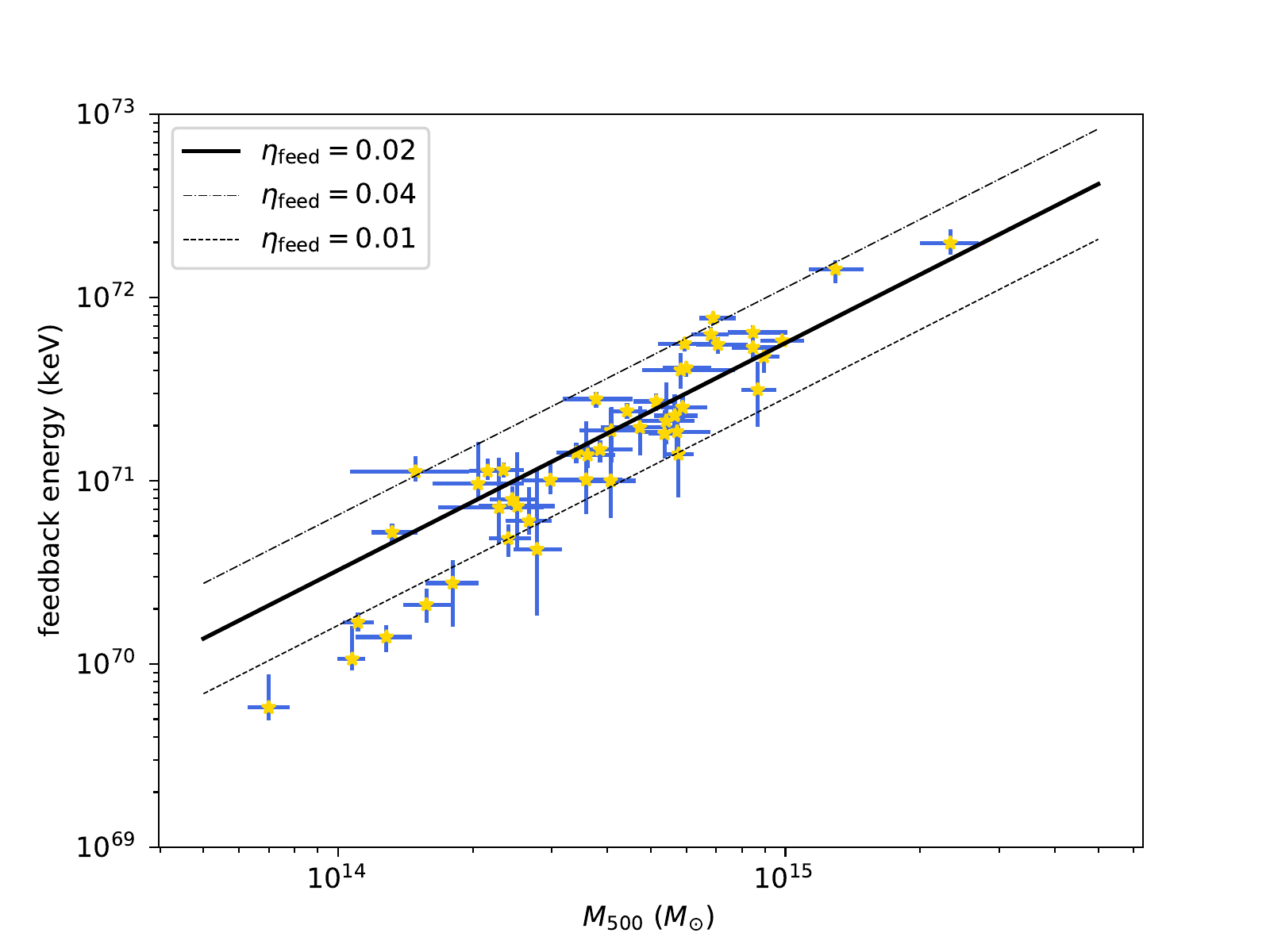}
	\caption{Total feedback energy of clusters or groups in the sample. \label{fig:m500-feed}}
\end{figure}

The total feedback energy within $r_{500}$ is estimated by integrating the $\Delta E_{\rm feed}(r)$
\begin{equation}
E_{\rm feed,tot}=\int^{r_{500}}_{0}  4\pi n_{\rm g}r^2\Delta E_{\rm feed}(r)/n^* dr.
\end{equation}
We plot and show the dependence of $E_{\rm feed,tot}$ on cluster mass in Figure \ref{fig:m500-feed}, in which the distribution of data roughly follows a power-law form. Assuming that the total feedback energy is fully provided by the SMBH sitting in the BCG, we estimate the feedback efficiency $\eta_{\rm feed}$, which is defined as the ratio of feedback energy to the energy corresponding to the rest mass of SMBH ($M_{\rm BH}$). By adopting the $M_{500}$-$M_{\rm BH}$ relation, derived from a sample of 71 galaxy clusters by \citet{phipps19}
\begin{equation}
\log_{10}\left(\frac{M_{\rm BH}}{10^9M_{\odot}}\right)=-0.82+1.16\log_{10}\left(\frac{M_{500}}{10^{13}M_{\odot}}\right),
\end{equation}
we obtained $\eta_{\rm feed}$ $\sim 0.02$ (Figure \ref{fig:m500-feed}), implying that the black holes may be non-spinning (e.g., \citealt{fabian12}). As a comparison, the feedback efficiency is usually assumed to be in the range of $0.005$ to $0.2$ in simulations (e.g., \citealt{sijacki07}; \citealt{puchwein08}; \citealt{sijacki08}). The estimation of SMBH feedback efficiency will be an upper limit if there exist heating sources other than the SMBHs (e.g., the pre-heating of gas before it accelerates into the galaxy clusters or groups), or possible mergers which would increase the entropy.

\section{Summary} \label{sec:summary}
We have investigated the entropy profiles of ICM in a sample of 47 galaxy clusters and groups that have been observed out to at least $\sim r_{500}$ using a physical model, which takes into account the effect of gravitational heating, work done via gas compression, net heat change through non-gravitational processes, non-thermal pressure, and the gas clumping. Our model has achieved acceptable fits to all of the sample members, and the best-fit sample-averaged ICM entropy profile is consistent with the power-law prediction from adiabatic simulations near the virial radius. The sample-averaged feedback energy profile derived from the best-fit entropy profile is consistent with zero at the $68\%$ confidence level outside $\sim 0.35$ $r_{200}$. Based on the relation of $M_{500}$ and the total feedback energy, we suggest that the upper limit of the feedback efficiency is $\sim 0.02$ for the SMBH of the BCG, which lies in the range of $0.005$ to $0.2$ that is usually used in cosmological simulations.


\acknowledgments
This work is supported by
the Ministry of Science and Technology of China
(grant Nos. 2018YFA0404601),
and the National Natural Science Foundation of China
(grant Nos. 11973033, 11835009, 11621303).


\begin{thebibliography}{}
\bibitem[Aho et al.(2014)]{aho14} Aho, K., Derryberry, D., Peterson, T.\ 2014, Ecology, 95, 631
\bibitem[Akaike(1974)]{akai74} Akaike, H.\ 1974, IEEE Transactions on Automatic Control, 19, 716
\bibitem[Akamatsu et al.(2011)]{akama11} Akamatsu, H., Hoshino, A., Ishisaki, Y., et al.\ 2011, \pasj, 63, S1019 
\bibitem[Akamatsu et al.(2017)]{akama17} Akamatsu, H., Mizuno, M., Ota, N., et al.\ 2017, \aap, 600, A100
\bibitem[Allen et al.(2001)]{allen01} Allen, S.~W., Schmidt, R.~W., \& Fabian, A.~C.\ 2001, \mnras, 328, L37 
\bibitem[Andreon \& Hurn(2013)]{andreon13} Andreon, S., \& Hurn, M.~A.\ 2012, Statistical Analysis and Data Mining: The ASA Data Science Journal, 6(1), 15-33.
\bibitem[Bautz et al.(2009)]{bautz09} Bautz, M.~W., Miller, E.~D., Sanders, J.~S., et al.\ 2009, \pasj, 61, 1117 
\bibitem[Blanton et al.(2010)]{blanton10} Blanton, E.~L., Clarke, T.~E., Sarazin, C.~L., et al.\ 2010, Proceedings of the National Academy of Science, 107, 7174
\bibitem[Buote et al.(2016)]{buote16} Buote, D.~A., Su, Y., Gastaldello, F., et al.\ 2016, \apj, 826, 146
\bibitem[Burnham \& Anderson(2002)]{burnham02} Burnham, K. P., Anderson, D. R.\ 2002, Model Selection and Multimodel Inference: A practical information-theoretic approach (2nd ed.), Springer-Verlag
%
\bibitem[Cavagnolo et al.(2009)]{cava09} Cavagnolo, K.~W., Donahue, M., Voit, G.~M., \& Sun, M.\ 2009, \apjs, 182, 12 
\bibitem[Chaudhuri et al.(2012)]{cnm12} Chaudhuri, A., Nath, B.~B., \& Majumdar, S.\ 2012, \apj, 759, 87
\bibitem[Churazov et al.(2012)]{chura12} Churazov, E., Vikhlinin, A., Zhuravleva, I., et al.\ 2012, \mnras, 421, 1123
\bibitem[Claeskens(2016)]{claeskens16} Claeskens, G.\ 2016, Annual Review of Statistics and Its Application, 3, 233
\bibitem[Eckert et al.(2019)]{eckert19} Eckert, D., Ghirardini, V., Ettori, S., et al.\ 2019, \aap, 621, A40 
\bibitem[Eckert et al.(2015)]{eckert15} Eckert, D., Roncarelli, M., Ettori, S., et al.\ 2015, \mnras, 447, 2198
\bibitem[Eckert et al.(2012)]{eckert12} Eckert, D., Vazza, F., Ettori, S., et al.\ 2012, \aap, 541, A57 
\bibitem[Engeland \& Gottschalk(2002)]{engeland02} Engeland, K., \& Gottschalk, L.\ 2002, Hydrology and Earth System Sciences, 6, 883
\bibitem[Fabian(2012)]{fabian12} Fabian, A.~C.\ 2012, \araa, 50, 455
\bibitem[Ghirardini et al.(2019)]{ghirar18} Ghirardini, V., Eckert, D., Ettori, S., et al.\ 2019, \aap, 621, A41.
\bibitem[Gitti et al.(2012)]{gitti12} Gitti, M., Brighenti, F., \& McNamara, B.~R.\ 2012, Advances in Astronomy, 2012, 950641
\bibitem[Grevesse \& Sauval(1998)]{grevesse98} Grevesse, N., \& Sauval, A.~J.\ 1998, \ssr, 85, 161 
\bibitem[Gu et al.(2016)]{gu16} Gu, L., Wen, Z., Gandhi, P., et al.\ 2016, \apj, 826, 72
\bibitem[Hastings(1970)]{Bimka70} Hastings, W. K., 1970, Biometrika, 57, 97
\bibitem[Hoshino et al.(2010)]{hoshino10} Hoshino, A., Henry, J.~P., Sato, K., et al.\ 2010, \pasj, 62, 371
\bibitem[Hudson et al.(2010)]{hudson10} Hudson, D.~S., Mittal, R., Reiprich, T.~H., et al.\ 2010, \aap, 513, A37
\bibitem[Ichikawa et al.(2013)]{ichikawa13} Ichikawa, K., Matsushita, K., Okabe, N., et al.\ 2013, \apj, 766, 90 
\bibitem[Iqbal et al.(2017)]{iqbal17} Iqbal, A., Kale, R., Majumdar, S., et al.\ 2017, Journal of Astrophysics and Astronomy, 38, 68
\bibitem[Kaiser(1986)]{kaiser86} Kaiser, N.\ 1986, \mnras, 222, 323 
\bibitem[Kalberla et al.(2005)]{kalberla05} Kalberla, P.~M.~W., Burton, W.~B., Hartmann, D., et al.\ 2005, \aap, 440, 775 
\bibitem[Kardar(2007)]{kardar07} Kardar, M.\ 2007, Statistical Physics of Particles, Cambridge University Press
\bibitem[Kawaharada et al.(2010)]{kawa10} Kawaharada, M., Okabe, N., Umetsu, K., et al.\ 2010, \apj, 714, 423 
\bibitem[Kettula et al.(2013)]{kett13} Kettula, K., Nevalainen, J., \& Miller, E.~D.\ 2013, \aap, 552, A47.
\bibitem[Khatri \& Gaspari(2016)]{khatri16} Khatri, R., \& Gaspari, M.\ 2016, \mnras, 463, 655 
\bibitem[Knuth(1997)]{knuth97} Knuth, D. E.\ 1997, The art of computer programming, volume 2 (3rd ed.): seminumerical algorithms, Addison-Wesley Longman Publishing Co. Inc.
\bibitem[Kotov \& Vikhlinin(2005)]{kotov05} Kotov, O., \& Vikhlinin, A.\ 2005, \apj, 633, 781
\bibitem[Kushino et al.(2002)]{kushino02} Kushino, A., Ishisaki, Y., Morita, U., et al.\ 2002, \pasj, 54, 327
\bibitem[Lakhchaura et al.(2016)]{lak16} Lakhchaura, K., Saini, T.~D., \& Sharma, P.\ 2016, \mnras, 460, 2625 
\bibitem[Lapi et al.(2010)]{lapi10} Lapi, A., Fusco-Femiano, R., \& Cavaliere, A.\ 2010, \aap, 516, A34
\bibitem[Lovell et al.(2018)]{lovell18} Lovell, M.~R., Pillepich, A., Genel, S., et al.\ 2018, \mnras, 481, 1950
\bibitem[Lovisari \& Reiprich(2019)]{lovisari19} Lovisari, L., \& Reiprich, T.~H.\ 2019, \mnras, 483, 540
\bibitem[Makishima et al.(2001)]{makishima01} Makishima, K., Ezawa, H., Fukuzawa, Y., et al.\ 2001, \pasj, 53, 401
\bibitem[Matsushita et al.(2003)]{matsushita03} Matsushita, K., Finoguenov, A., \& B{\"o}hringer, H.\ 2003, \aap, 401, 443
\bibitem[Mazzotta et al.(2004)]{mazza04} Mazzotta, P., Rasia, E., Moscardini, L., et al.\ 2004, \mnras, 354, 10
\bibitem[Mernier et al.(2019)]{mernier19} Mernier, F., Werner, N., Lakhchaura, K., et al.\ 2019, arXiv e-prints, arXiv:1911.09684 
\bibitem[Miller et al.(2012)]{miller12} Miller, E.~D., Bautz, M., George, J., et al.\ 2012, American Institute of Physics Conference Series, 1427, 13
\bibitem[Morandi \& Cui(2014)]{morandi14} Morandi, A., \& Cui, W.\ 2014, \mnras, 437, 1909
\bibitem[Morandi \& Sun(2016)]{morandi16} Morandi, A., \& Sun, M.\ 2016, \mnras, 457, 3266 
\bibitem[N{\ae}ss(2012)]{naess12} N{\ae}ss, S.~K.\ 2012, \aap, 538, A17
\bibitem[Nagai \& Lau(2011)]{nagai11} Nagai, D., \& Lau, E.~T.\ 2011, \apjl, 731, L10 \
\bibitem[Nakajima et al.(2018)]{nakajima18} Nakajima, H., Maeda, Y., Uchida, H., et al.\ 2018, \pasj, 70, 21
\bibitem[Nash \& Sutcliffe(1970)]{nash70} Nash, J.~E., \& Sutcliffe, J.~V.\ 1970, Journal of Hydrology, 10, 282 
\bibitem[Navarro et al.(1997)]{nfw97} Navarro, J.~F., Frenk, C.~S., \& White, S.~D.~M.\ 1997, \apj, 490, 493 
\bibitem[Nelson et al.(2014)]{nel14} Nelson, K., Lau, E.~T., \& Nagai, D.\ 2014, \apj, 792, 25 
\bibitem[Okabe et al.(2014)]{okabe14} Okabe, N., Umetsu, K., Tamura, T., et al.\ 2014, \pasj, 66, 99
\bibitem[Ostriker et al.(2005)]{ostriker05} Ostriker, J.~P., Bode, P., \& Babul, A.\ 2005, \apj, 634, 964
\bibitem[Panagoulia et al.(2014)]{panagoulia14} Panagoulia, E.~K., Fabian, A.~C., \& Sanders, J.~S.\ 2014, \mnras, 438, 2341 
\bibitem[Patej \& Loeb(2015)]{patej15} Patej, A., \& Loeb, A.\ 2015, \apjl, 798, L20
\bibitem[Phipps et al.(2019)]{phipps19} Phipps, F., Bogd{\'a}n, {\'A}., Lovisari, L., et al.\ 2019, \apj, 875, 141
\bibitem[Planelles et al.(2013)]{planelles13} Planelles, S., Borgani, S., Dolag, K., et al.\ 2013, \mnras, 431, 1487
\bibitem[Pointecouteau et al.(2004)]{point04} Pointecouteau, E., Arnaud, M., Kaastra, J., et al.\ 2004, \aap, 423, 33
\bibitem[Press \& Schechter(1974)]{ps74} Press, W.~H., \& Schechter, P.\ 1974, \apj, 187, 425
\bibitem[Puchwein et al.(2008)]{puchwein08} Puchwein, E., Sijacki, D., \& Springel, V.\ 2008, \apjl, 687, L53
\bibitem[Reiprich et al.(2009)]{reiprich09} Reiprich, T.~H., Hudson, D.~S., Zhang, Y.-Y., et al.\ 2009, \aap, 501, 899 
\bibitem[Roncarelli et al.(2006)]{ron06} Roncarelli, M., Ettori, S., Dolag, K., et al.\ 2006, \mnras, 373, 1339
\bibitem[Sato et al.(2014a)]{sato14a} Sato, K., Matsushita, K., Tamura, T., et al.\ 2014, Suzaku-maxi 2014: Expanding the Frontiers of the X-ray Universe, 414
\bibitem[Sato et al.(2014b)]{sato14b} Sato, K., Matsushita, K., Yamasaki, N.~Y., et al.\ 2014, \pasj, 66, 85
\bibitem[Sato et al.(2011)]{sato11} Sato, T., Matsushita, K., Ota, N., et al.\ 2011, \pasj, 63, S991 
\bibitem[Sato et al.(2012)]{sato12} Sato, T., Sasaki, T., Matsushita, K., et al.\ 2012, \pasj, 64, 95 
\bibitem[Schellenberger et al.(2015)]{S15} Schellenberger, G., Reiprich, T.~H., Lovisari, L., Nevalainen, J., \& David, L.\ 2015, \aap, 575, A30 
\bibitem[Sijacki et al.(2008)]{sijacki08} Sijacki, D., Pfrommer, C., Springel, V., et al.\ 2008, \mnras, 387, 1403
\bibitem[Sijacki \& Springel(2007)]{sijacki07} Sijacki, D., \& Springel, V.\ 2007, Heating Versus Cooling in Galaxies and Clusters of Galaxies, 237
\bibitem[Simionescu et al.(2011)]{simon11} Simionescu, A., Allen, S.~W., Mantz, A., et al.\ 2011, Science, 331, 1576 
\bibitem[Simionescu et al.(2017)]{simion17} Simionescu, A., Werner, N., Mantz, A., et al.\ 2017, \mnras, 469, 1476
\bibitem[Simionescu et al.(2013)]{simion13} Simionescu, A., Werner, N., Urban, O., et al.\ 2013, \apj, 775, 4
\bibitem[Springel et al.(2018)]{TNG18} Springel, V., Pakmor, R., Pillepich, A., et al.\ 2018, \mnras, 475, 676
\bibitem[Snowden et al.(2008)]{snowden08} Snowden, S.~L., Mushotzky, R.~F., Kuntz, K.~D., et al.\ 2008, \aap, 478, 615
\bibitem[Su et al.(2015)]{su15} Su, Y., Buote, D., Gastaldello, F., \& Brighenti, F.\ 2015, \apj, 805, 104 
\bibitem[Su et al.(2013)]{su13} Su, Y., White, R.~E., \& Miller, E.~D.\ 2013, \apj, 775, 89
\bibitem[Sugizaki et al.(2009)]{sugizaki09} Sugizaki, M., Kamae, T., \& Maeda, Y.\ 2009, \pasj, 61, S55
\bibitem[Szyd{\l}owski et al.(2015)]{szy15} Szyd{\l}owski, M., Krawiec, A., Kurek, A., et al.\ 2015, European Physical Journal C, 75, 5
\bibitem[Tchernin et al.(2016)]{tcher16} Tchernin, C., Eckert, D., Ettori, S., et al.\ 2016, \aap, 595, A42.
\bibitem[Th{\"o}lken et al.(2016)]{tholken16} Th{\"o}lken, S., Lovisari, L., Reiprich, T.~H., et al.\ 2016, \aap, 592, A37
\bibitem[Tozzi \& Norman(2001)]{tozzi01} Tozzi, P., \& Norman, C.\ 2001, \apj, 546, 63 
\bibitem[Trotta(2017)]{trotta17} Trotta, R.\ 2017, arXiv e-prints, arXiv:1701.01467
\bibitem[Truong et al.(2018)]{truong18} Truong, N., Rasia, E., Mazzotta, P., et al.\ 2018, \mnras, 474, 4089
\bibitem[Urban et al.(2014)]{urban14} Urban, O., Simionescu, A., Werner, N., et al.\ 2014, \mnras, 437, 3939
\bibitem[van der Vaart(1998)]{vaart98} van der Vaart, A.W.\ 1998, Asymptotic statistics, Cambridge University Press
\bibitem[Vazza(2011)]{vazza11} Vazza, F.\ 2011, \mnras, 410, 461
\bibitem[Vazza et al.(2013)]{vazza13} Vazza, F., Eckert, D., Simionescu, A., Br{\"u}ggen, M., \& Ettori, S.\ 2013, \mnras, 429, 799 
\bibitem[Vikhlinin et al.(2006)]{vikhlinin06} Vikhlinin, A., Kravtsov, A., Forman, W., et al.\ 2006, \apj, 640, 691
\bibitem[Vikhlinin et al.(2005)]{vikhlinin05} Vikhlinin, A., Markevitch, M., Murray, S.~S., et al.\ 2005, \apj, 628, 655 
\bibitem[Voit et al.(2003)]{voit03} Voit, G.~M., Balogh, M.~L., Bower, R.~G., et al.\ 2003, \apj, 593, 272
\bibitem[Voit et al.(2002)]{voit02} Voit, G.~M., Bryan, G.~L., Balogh, M.~L., \& Bower, R.~G.\ 2002, \apj, 576, 601 
\bibitem[Voit et al.(2005)]{voit05b} Voit, G.~M., Kay, S.~T., \& Bryan, G.~L.\ 2005, \mnras, 364, 909 
\bibitem[Walker et al.(2013)]{walker13} Walker, S.~A., Fabian, A.~C., Sanders, J.~S., et al.\ 2013, \mnras, 432, 554 
\bibitem[Walker et al.(2012a)]{walker12a} Walker, S.~A., Fabian, A.~C., Sanders, J.~S., \& George, M.~R.\ 2012a, \mnras, 427, L45
\bibitem[Walker et al.(2012b)]{walker12c} Walker, S.~A., Fabian, A.~C., Sanders, J.~S., \& George, M.~R.\ 2012b, \mnras, 424, 1826 
\bibitem[Walker et al.(2012c)]{walker12b} Walker, S.~A., Fabian, A.~C., Sanders, J.~S., George, M.~R., \& Tawara, Y.\ 2012c, \mnras, 422, 3503 
\bibitem[Wong \& Sarazin(2009)]{wong09} Wong, K.-W., \& Sarazin, C.~L.\ 2009, \apj, 707, 1141
\bibitem[Zhang et al.(2006)]{zhang06} Zhang, Y.-Y., B{\"o}hringer, H., Finoguenov, A., et al.\ 2006, \aap, 456, 55 
\bibitem[Zhu et al.(2016)]{zhu16} Zhu, Z., Xu, H., Wang, J., et al.\ 2016, \apj, 816, 54 
\bibitem[Zhuravleva et al.(2015)]{zhurav15} Zhuravleva, I., Churazov, E., Ar{\'e}valo, P., et al.\ 2015, \mnras, 450, 4184

\end{thebibliography}
\end{document}